\documentclass[sigconf]{acmart}

\AtBeginDocument{%
  }

\copyrightyear{2025}
\acmYear{2025}
\setcopyright{cc}
\setcctype{by-sa}
\acmConference[DIS '25]{Designing Interactive Systems Conference}{July 5--9, 2025}{Funchal, Portugal}
\acmBooktitle{Designing Interactive Systems Conference (DIS '25), July 5--9, 2025, Funchal, Portugal}\acmDOI{10.1145/3715336.3735700}
\acmISBN{979-8-4007-1485-6/2025/07}

\usepackage{transparent}
\usepackage{subcaption}
\usepackage{tabularx}
\usepackage{longtable} % tables that can span several pages
\usepackage{colortbl}
\usepackage{multirow}
\usepackage{rotating}
\usepackage{enumitem}
\usepackage{listings}

\begin{document}

%%
%% The "title" command has an optional parameter,
%% allowing the author to define a "short title" to be used in page headers.
\title{Exploring Anthropomorphism in Conversational Agents for Environmental Sustainability}

%%
%% The "author" command and its associated commands are used to define
%% the authors and their affiliations.
%% Of note is the shared affiliation of the first two authors, and the
%% "authornote" and "authornotemark" commands
%% used to denote shared contribution to the research.
\author{Mathyas Giudici}
\orcid{0000-0002-4935-5131}
\affiliation{%
  \institution{Politecnico di Milano}
  \city{Milan}
  \country{Italy}
}
\email{mathyas.giudici@polimi.it}

\author{Samuele Scherini}
\orcid{0009-0001-5520-3804}
\affiliation{%
  \institution{Politecnico di Milano}
  \city{Milan}
  \country{Italy}
}
\email{samuele.scherini@mail.polimi.it}

\author{Pascal Chaussumier}
\orcid{0009-0003-9080-7499}
\affiliation{%
  \institution{EDF - Électricité de France}
  \city{Paris}
  \country{France}
}
\affiliation{%
  \institution{Edison}
  \city{Milan}
  \country{Italy}
}
\email{pascal.chaussumier@edison.it}

\author{Stefano Ginocchio}
\orcid{0000-0003-0952-6212}
\affiliation{%
  \institution{Edison}
  \city{Milan}
  \country{Italy}
}
\email{stefano.ginocchio@edison.it}

\author{Franca Garzotto}
\orcid{0000-0003-4905-7166}
\affiliation{%
  \institution{Università degli Studi di Milano-Bicocca}
  \city{Milan}
  \country{Italy}
}
\affiliation{%
  \institution{Politecnico di Milano}
  \city{Milan}
  \country{Italy}
}
\email{franca.garzotto@polimi.it}
%%
%% By default, the full list of authors will be used in the page
%% headers. Often, this list is too long, and will overlap
%% other information printed in the page headers. This command allows
%% the author to define a more concise list
%% of authors' names for this purpose.
\renewcommand{\shortauthors}{Giudici and Scherini et al.}

%%
%% The abstract is a short summary of the work to be presented in the
%% article.
\begin{abstract}
The paper investigates the integration of Large Language Models (LLMs) into Conversational Agents (CAs) to encourage a shift in consumption patterns from a demand-driven to a supply-based paradigm. Specifically, the research examines the role of anthropomorphic design in delivering environmentally conscious messages by comparing two CA designs: a personified agent representing an appliance and a traditional, non-personified assistant. A lab study (N=26) assessed the impact of these designs on interaction, perceived self-efficacy, and engagement. Results indicate that LLM-based CAs significantly enhance users' self-reported eco-friendly behaviors, with participants expressing greater confidence in managing energy consumption. While the anthropomorphic design did not notably affect self-efficacy, those interacting with the personified agent reported a stronger sense of connection with the system. These findings suggest that although anthropomorphic CAs may improve user engagement, both designs hold promise for fostering sustainable behaviors in home energy management.
\end{abstract}

%%
%% The code below is generated by the tool at http://dl.acm.org/ccs.cfm.
%% Please copy and paste the code instead of the example below.
%%


%%
%% Keywords. The author(s) should pick words that accurately describe
%% the work being presented. Separate the keywords with commas.


% \received{20 February 2007}
% \received[revised]{12 March 2009}
% \received[accepted]{5 June 2009}

%%
%% This command processes the author and affiliation and title
%% information and builds the first part of the formatted document.
\maketitle

\section{Introduction}
\label{sec:introduction}
Within the current climate change context, residential energy consumption contributes significantly to global emissions~\cite{EuropeanCom2006}. Sustainable Human-Computer Interaction (S-HCI) \cite{disalvo2010mapping} has emerged as an important field that investigates the design and impact of systems to encourage environmentally friendly behaviors among individuals and communities~\cite{costanzaDoingLaundryAgents2014, alan_tariff_2016, panagiotidouSolarClubSupportingRenewable2024}.
Among the tools explored in S-HCI, Conversational Agents (CAs) -- digital tools that empower people to interact with technology using natural language \cite{hussain2019survey} -- have gained traction across various domains~\cite{diederich2019promoting, grassini2024systematic}.
In particular, in the environmental sustainability field, they are a promising tool to promote sustainable practices~\cite{fontechaUsingConversationalAssistants2019,giudici2022candy}. These systems can assist users in adopting environmentally friendly behaviors, such as supply-based energy management, which aligns energy consumption with periods when renewable resources, like solar or wind power, are most available~\cite{panagiotidouSolarClubSupportingRenewable2024}.

Building on this progress, researchers have recently turned their attention to a subset of CAs powered by Large Language Models (LLMs) \cite{vaswani2017attention}, which brings notable potential to sustainability efforts. LLMs extend traditional CAs with advanced language processing that engages users in more natural and interactive conversations. Specifically, LLM-powered chatbots have been adopted to investigate their effectiveness in delivering eco-friendly messages and encouraging sustainable habits~\cite{king2023sasha, designs8030043,giudici2025generatinghomeassistantautomationsusing}.

However, the role of \textit{anthropomorphization} in such systems has not yet been fully understood. We refer to anthropomorphization, also known as personification, as the process of giving inanimate or non-human entities human characteristics, attributes, and emotions \cite{damianoLuisaAnthropomorphism, epleyOnSeeingAHuman, kimAnthropomorphismComputersIt2012, zlotowskiAnthropomorphismOpportunities}.
The effects of anthropomorphic design in LLM-based agents must be investigated to gather insights into users’ perceptions of such systems. Understanding this peculiar aspect could provide valuable insights into the impact of emotional and relatable design on user engagement and behavior \cite{yuan_emotion_2013,meiselwitz_experimental_2020}.

Leveraging on the work of \citet{costanzaDoingLaundryAgents2014}, we present \emph{Washy}, an LLM-powered CA system fully integrated with a smart home eco-system, specifically our lab located in \textit{Officine Edison}'s Smart Home laboratory (Milan, Italy).
\emph{Washy} includes two distinct agents -- the first with a higher level of anthropomorphism that personifies the appliance and the second with a more neutral persona acting as a traditional assistant -- allowing us to investigate how varying degrees of agency influence user interaction. The system enables users to schedule their laundry cycles based on an estimated forecast of their solar panel energy production and remotely controls the laboratory washing machine.

Relying on the \emph{Washy} system, we designed and conducted a pilot study involving 26 participants who were divided into two groups, each interacting with one of the two agents. This setup allowed us to evaluate the system's effectiveness in promoting eco-friendly behaviors, as well as the effect of anthropomorphism on user interaction and engagement.

Our findings suggest that LLM-based CAs can positively influence users’ engagement with sustainable practices. The results indicate that user interactions with \emph{Washy} increased self-reported eco-conscious behaviors, while the impact of anthropomorphism and the preference between traditional and personified agents varied without being able to make a definitive claim. Nevertheless, this manuscript contributes to Sustainable HCI by advancing the understanding of how LLM-powered CAs and their anthropomorphization affect environmental behaviors in a domestic setting and brings preliminary results for future work in the field. 

The paper is structured as follows. Section~\ref{sec:rel-works} covers the current state-of-the-art in S-HCI, focusing on the use of conversational technologies and Large Language Models for promoting environmental sustainability. Section~\ref{sec:system}  describes the \emph{Washy} system, focusing on the design and the implementation of the Personified and Traditional assistants.
Section~\ref{sec:empiricalstud} explains the methodology of our study with a detailed description of each evaluation metric. Section~\ref{sec:results} presents the collected results, while Section~\ref{sec:discussion} discusses the findings and their implication. Section~\ref{sec:limitations} outlines the limitations, while Section~\ref{sec:conclusions} concludes and sum up the work.

\section{Related Work}
\label{sec:rel-works}

\subsection{Sustainable HCI and Eco-Feedback}

Climate concerns and sustainability are critical challenges in addressing global warming and its impacts. The Intergovernmental Panel on Climate Change claims that human activity is the primary cause of climate change~\cite{masson2021climate}. Despite not being the only factor, as other contributors like industrial emissions and deforestation also play significant roles~\cite{intergovernmentalpanelonclimatechangeipccClimateChange20222023}, individual energy consumption contributes to this problem. In particular, residential electricity consumption accounts for a significant portion of the world's energy use~\cite{iea2022world}.

To advance decarbonization and mitigate climate change, renewable energy sources must be adopted, along with an understanding of their nuances and technical intricacies. Mastering the complexities of current renewable technologies is key to successfully transitioning from fossil fuels to a more sustainable energy system~\cite{osmanCostEnvironmentalImpact2023}.

The shift from a demand-driven to a supply-based consumption model is critical for a successful switch to renewable energy, where energy use is matched with the availability of renewable sources such as wind or solar energy~\cite{EuropeanCom2006}. Although it can be beneficial, storing the generated energy in batteries comes with costs and environmental effects~\cite{costanza2012understanding, EuropeanCom2006}. Demand-shifting, which involves scheduling energy-intensive activities for periods when renewable energy is available, offers a more cost-effective and ecological solution~\cite{pierceLocalEnergyIndicator2012}.

Research in Human-Computer Interaction (HCI) can play an essential role in creating new digital solutions that can be adopted in our daily lives to raise awareness and guide individuals toward more sustainable behaviors~\cite{vinuesa2020role,disalvo2010mapping}. Studies in this field have shown that digital systems providing sustainability and environmental feedback, such as real-time visualizations of energy consumption, encourage more mindful electricity use \cite{al2022interactive,costanza2012understanding}.

Many of these studies involve the development and analysis of monitoring systems, which track household energy consumption and provide feedback to users. These systems focus on raising awareness by presenting data about energy use, while leaving the actual decision-making and actions to the householder. Unlike automatic managers, which actively control how devices operate, monitors rely on feedback to inform users and encourage voluntary behavior change~\cite{van_dam2010home, jensenAssistedShiftingElectricity2018}.

Monitors in home energy systems play a crucial role in promoting sustainable practices. They help householders act more sustainably by tracking energy consumption and providing feedback that raises awareness and fosters behavioral changes. A widely used approach within this category is eco-feedback, which visualizes past and present energy consumption to prompt users to adopt more sustainable habits~\cite{froehlich2010design, kluckner2013exploring}. Rooted in behavioral psychology, eco-feedback operates on the assumption that providing the right information will encourage people to modify their behavior to conserve energy.

One first notable example is the literature provided by Costanza et al.~\cite{costanzaDoingLaundryAgents2014}, who proposed ``Agent B'', an interactive, agent-based system capable of predicting the price of washing machine usage based on weather forecast data, which notifies the user if the price rises more than a certain user-defined threshold.
Similarly, Alan et al.~\cite{alan_tariff_2016} developed a home energy management system called Tariff, which monitors household energy consumption, as well as available energy tariffs. The purpose of ``Tariff'' is to evaluate how users would perceive an autonomous tariff-switching system that may affect their economic situation. The study also investigated the level of control that users would delegate to the system.
While studying the influence of social feedback, Ham et al.~\cite{hamRobotThatSays2009} found that the effect of social feedback on energy conservation is the joint effect of negative and positive feedback. Although a system that offers compliments may be pleasing and enjoyable for users, it does not necessarily ensure a lasting change in their behavior.

Despite over a decade of research into persuasive technology within Sustainable HCI, most eco-feedback systems have primarily been designed using visual interfaces~\cite{costanza2012understanding,pierceLocalEnergyIndicator2012,alan_tariff_2016}. Physical interaction approaches that directly engage users have also played a significant role~\cite{jensen2018washing,Morais2021,Quintal2016}.
% These visual-based methods dominate the HCI landscape by providing users with real-time feedback on their energy consumption.
Additionally, more recent studies have explored eco-forecasting, where predictions about future consumption, costs, or grid demand are presented, allowing householders to shift their energy use to more optimal times~\cite{costanzaDoingLaundryAgents2014, alan_tariff_2016}.
However, there is still much work to be done in exploring alternative interaction paradigms, such as conversational technologies \cite{sanguinetti2024conversational}, that could better support users in adopting more environmentally friendly behaviors.

\subsection{Conversational Technologies for Environmental Sustainability}
\label{sec:conv-related}
Conversational Agents (CA) are a class of software systems designed to engage in conversations with users via natural language processing. They can interpret and generate human-like responses enabling both text and speech interaction with the end user~\cite{hussain2019survey}.
In the past, rule-based chatbot established themselves as the standard approach for developing conversational agents. Rule-based chatbots operate by scanning user utterances and responding in accordance with a predefined set of rules. Under this paradigm, the first chatbot ELIZA~\cite{weizenbaum1966eliza} rudimentarily parsed user input for patterns and produced responses by using pre-programmed templates. This class of conversational agents still plays a significant role in different domains~\cite{SocialMedia, CustomerService}, largely facilitated by low-code tools (e.g.,  Google’s Dialogflow~\cite{sabharwalDialogFlow}).

In recent years, thanks to the recent advances in the Generative Artificial Intelligence (GenAI) field~\cite{GenerativeHistory}, chatbots are becoming more sophisticated. Leveraging the capabilities of modern LLMs (Large Language Models) \cite{vaswani2017attention}, they are able to generate replies using machine-learning techniques, based on the context of the conversation. This technology allows cutting-edge chatbots to produce dynamic responses, engaging the user in a more natural and human-like conversation.

Although LLMs have demonstrated remarkable capabilities in generating natural language text, they need to be carefully adopted to harness their full potential in specific use cases \cite{dam2024complete}.
The performance of LLMs is highly sensitive to the input prompt. Thus, the effective usage of an emerging technique, called prompt engineering, is highly required to generate meaningful answers~\cite{PromptEngineering}. 
A well-designed prompt can produce performances that are comparable to those of a fine-tuned model~\cite{brown2020languagemodelsfewshotlearners}, opening up AI to a wider range of researchers and fields. % However, since LLMs are difficult to interpret, prompt engineering lacks a rigorous, standardized procedure; instead, it mostly depends on the user's intuition, which necessitates consulting previous research to find appropriate prompting norms.

In recent years, conversational agents have been used in many different fields such as healthcare systems~\cite{grassini2024systematic}, sustainable mobility~\cite{diederich2019promoting}, and energy consumption~\cite{giudici2024delivering, fontechaUsingConversationalAssistants2019}.

In particular, in 2018, \citet{sannon2018personification} explored how personification and interactivity in conversational agents influence users’ willingness to disclose stress-related information. In a study with 441 participants, they found that a personified chatbot elicited more chronic and finance-related stress disclosures but fewer home-related and detailed disclosures compared to a survey or non-personified chatbot. The results suggest that personification, more than interactivity, affects both the type and depth of sensitive disclosures, with implications for designing supportive conversational interfaces.

Still, \citet{diederich2019promoting} studied how to promote sustainable mobility using a conversational agent. The study found a positive impact of the perceived humanness of the conversational agent on perceived persuasiveness, indicating that the human-like design and human perception of such agents indeed contribute to persuasiveness.
Still, \citet{sNovelAIbasedChatbot2023} study highlights the possibility to revolutionize the way individuals interact with healthcare systems. By leveraging the OpenAI GPT-3.5 model, they developed a medical chatbot that can grant users insights and personalized diagnoses based on their symptoms.

Several studies have examined the broad use of conversational agents such as Alexa in smart homes, highlighting the ways in which this technology is incorporated into everyday life~\cite{sciuto2018hey}.
In addition, \cite{purington2017alexa} analyzed Amazon Echo user reviews to examine how people personify the device and engage in social interactions with its voice assistant (i.e., Alexa). They found that greater personification -- using Alexa’s name and personal pronouns -- correlates with higher sociability and increased user satisfaction, especially in multi-member households. Their findings suggest that conversational agents like Alexa are not only functional tools but also perceived social actors whose human-like traits influence how users integrate them into daily life.
Still, \citet{cho2019once} found that users naturally personified Alexa due to its human-like voice and conversational interface, often treating it as a friend or companion in the early stages. However, this personification led to unmet expectations, as users anticipated human-level emotional responses, relationship building, and conversational depth that Alexa could not fulfill. Over time, this mismatch caused disappointment and emotional detachment, ultimately reducing engagement and leading users to reframe Alexa as a mere tool or background object.
Lastly, \citet{reddy2021making} investigated how everyday household objects might express unique personalities and perspectives if endowed with voice, using speculative ``Thing Interviews'' to imagine deeply personalized interactions. By shifting from generic, function-driven voice assistants to more character-rich, context-aware things, the study emphasizes how personalization could emerge from an object’s materiality, history, and relationship with its human user.
The work highlights the potential of designing voice interfaces that are not only responsive but also reflective of a thing’s identity and role in domestic life, enriching the emotional and social fabric of human-object interactions.

Despite different opportunities and challenges to integrate conversational technologies in smart home environments, such digital tools have been demonstrated to be beneficial in supporting sustainable behavior~\cite{giudici2022candy}.
More specifically, there are widely recognized examples of conversational agents in research related to HCI and environmental sustainability that utilize eco-feedback.

In 2018, \citet{gnewuchDesigningConversationalAgents2018} implemented a conversational agent that provides energy feedback in a more innovative way, focusing less on visual feedback involving data. They successfully leveraged the chatbot's natural language capabilities to answer questions and provide personalized feedback. Ultimately, they suggest that future researchers investigate a combination of text-based and voice-based conversational agents.

\citet{fontechaUsingConversationalAssistants2019} developed a conversational agent focused on the interaction between users and their smart appliances. The chatbot is mainly speech-based, and it was designed with Google DialogFlow. The study aimed to investigate the interaction between the users and their devices.

Moreover, \citet{designs8030043} presented GreenIFTTT, a web-based conversational agent empowered by the GPT4 model, designed to encourage environmentally conscious habits within households. The system focuses on creating smart routines: automation sequences within the home environment based on the sequential execution of specific activities triggered by various conditions. They found out that integrating GPT4 into a conversational agent for promoting eco-friendly practices represents a significant step toward leveraging advanced AI technologies for environmental sustainability.

In 2023, \citet{giudici2023leafy} proposed ``Leafy'', a conversational agent embedded in a Smart Mirror to enhance home energy efficiency through gamification. The agent provides real-time feedback and daily challenges to encourage sustainable behavior and optimize energy consumption. The agent had a personification trait consisting of a tree-like avatar whose foliage changes according to the user's behaviors on a 5-level scale, starting from a few withered leaves to a lush foliage.

Similarly, \citet{berneyCareBasedEcoFeedbackAugmented2024} developed a generative AI-based chatbot to provide emotional eco-feedback to users. This study showed that the pathway from artifact to behavior was also effective for individuals who are less environmentally aware, which opens up a promising avenue to reach a demographic that would benefit the most from such interventions.

Still, this study suggests investigating the effects of emotional attachment as a motivational factor to mediate pro-environmental behavior.
While some studies suggest designing an eco-feedback system to provide both positive and negative social feedback~\cite{hamRobotThatSays2009}, it is still unclear if the degree of the agency may affect user engagement in a significant way~\cite{middenIllusionAgencyInfluence2012}. 

Inspired by the previous work of Costanza et al.~\cite{costanzaDoingLaundryAgents2014}, we developed an LLM-powered conversational agent aimed at helping users manage their laundry cycles. In this research, we try to highlight whether two different degrees of anthropomorphism of the system could influence users' responses to upcoming environmental challenges, such as energy efficiency and supply-based consumption.
To the best of our knowledge, no one has tried to investigate the influence of such personification of conversational agents in the context of individual demand-shift electricity behavioral change.

\section{System}
\label{sec:system}

\subsection{Design Rationale}
Drawing on prior work (see Section \ref{sec:rel-works}) and building directly on the laundry booking system developed by Costanza et al.~\cite{costanzaDoingLaundryAgents2014}, we propose \emph{Washy}, a probe of a conversational-based tool designed to enable users to plan their laundry activities according to the estimated power generated by their home solar panels. In order to maximize the energy generated by the solar panels and prevent the user from drawing power from the electrical grid when unnecessary, the system aims to recommend the optimal time for users to do their laundry, when solar energy generation is at its peak. 
The user interacts with a generative AI-powered conversational agent (GPT-4o) that offers custom recommendations. A smooth and energy-efficient laundry scheduling experience is made possible by these recommendations, which are tailored based on the user's solar panel and washing machine specifications.

\begin{figure}[ht]
    \centering
    \begin{subfigure}[t]{\linewidth}
        \includegraphics[width=\linewidth]{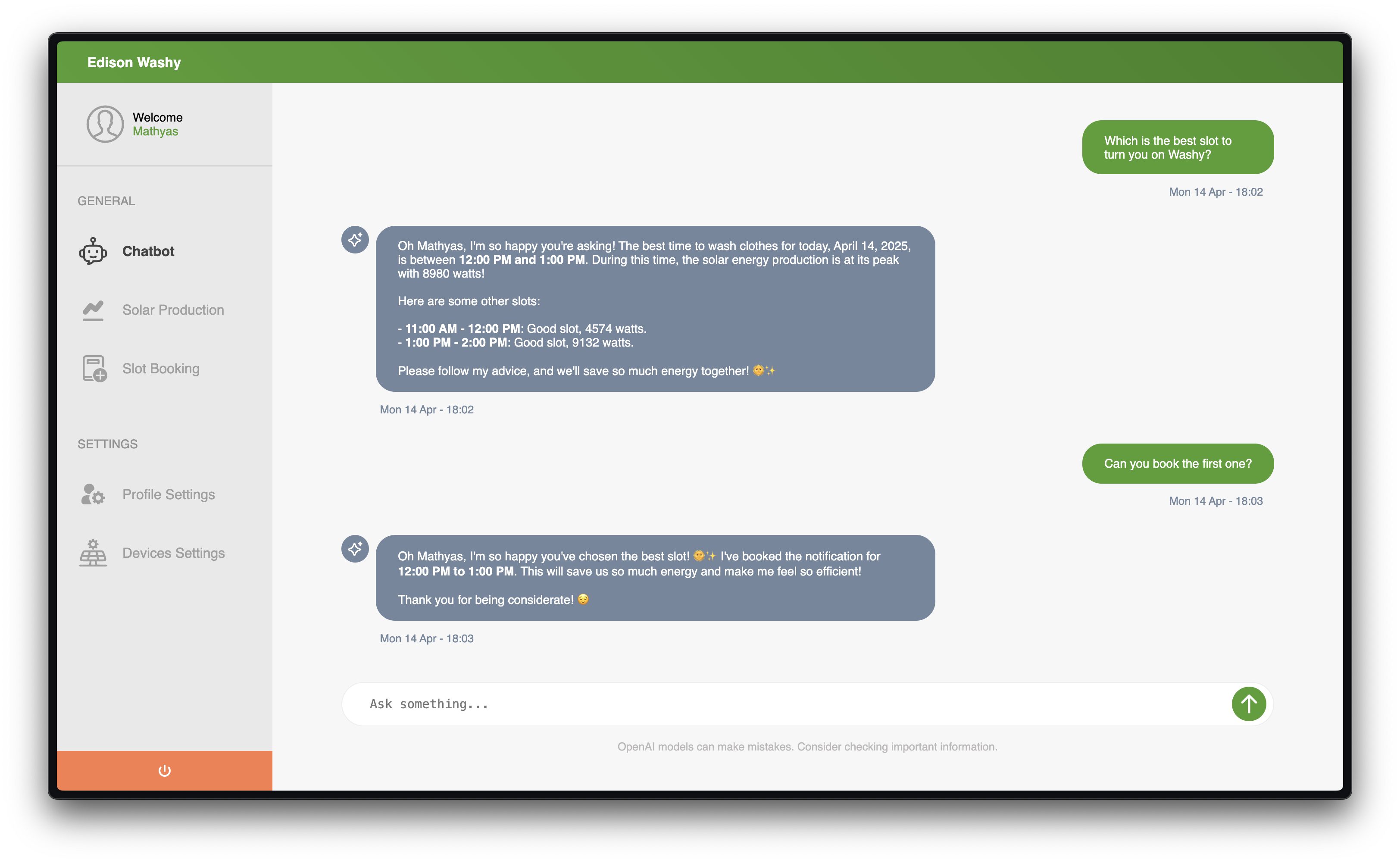}
        \caption{Example of a slot booking in the Chatbot Page}
        \label{fig:chat-booking}
    \end{subfigure}
    \hfill
    \begin{subfigure}[t]{\linewidth}
        \includegraphics[width=\linewidth]{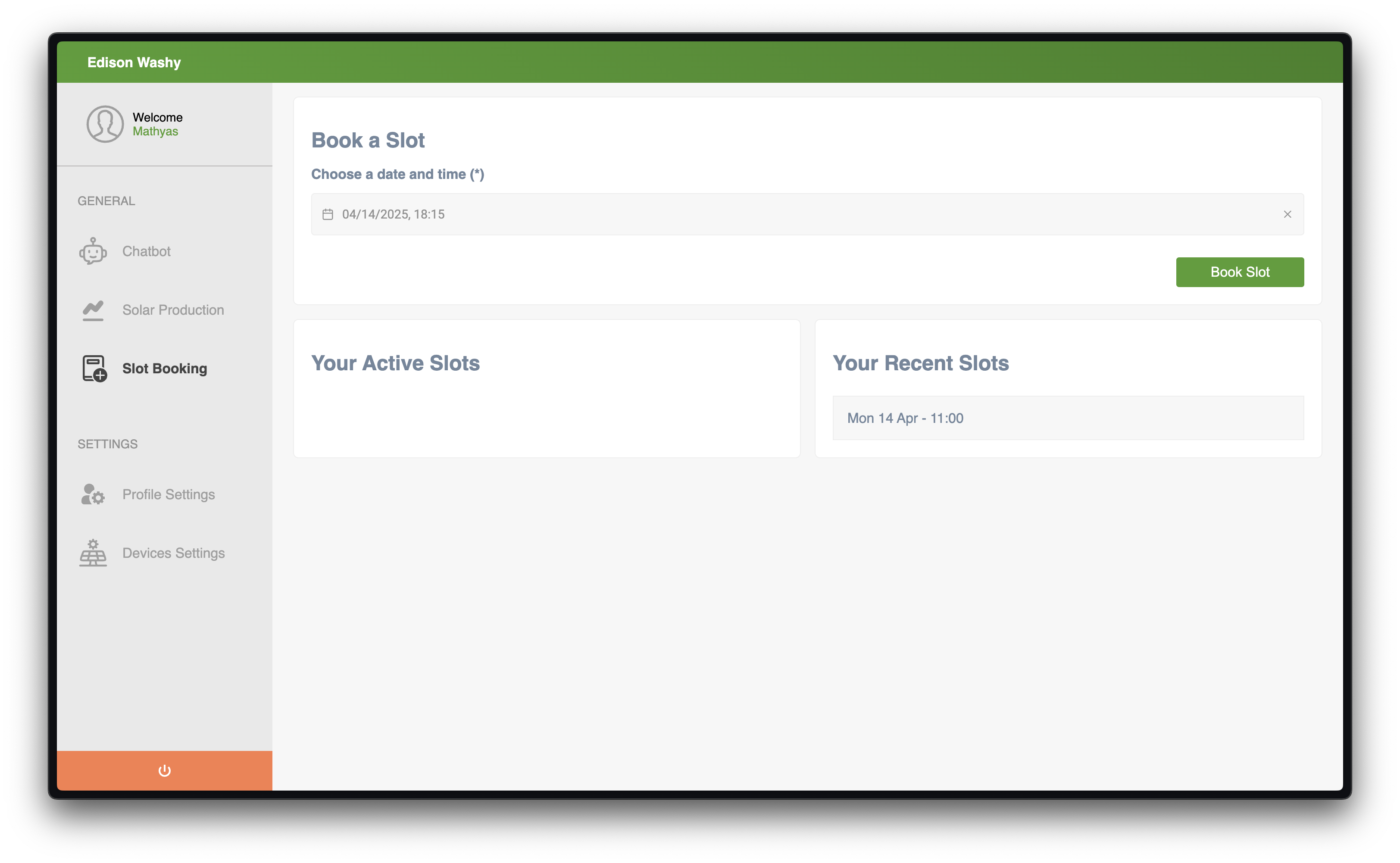}
        \caption{Upcoming slot booking page}
        \label{fig:ui-slots}
    \end{subfigure}
    \caption{Snapshot from the Application}
    \Description{Snapshot from the Application with the page of the chatbot demonstrating the booking of a slot, and the slot summary page.}
\end{figure}

Users ask the virtual assistant to suggest when they can run a laundry cycle for a given amount of time and -- if needed -- be able to specify the cycle's energy consumption. If no energy usage is defined, the system estimates using the washing machine's standard power consumption.
In order to provide the user with the best options based on solar energy availability, the virtual assistant evaluates and ranks all the time slots of the designated duration over the next few days. Then the user can ask the agent to set a reminder for that particular slot after choosing a time that works for them. The user is notified by the system when the reminder is about to happen, and they have the option to accept or reject the time of the laundry cycle. If the user confirms, the washing machine will start automatically at the time specified in the notification.
Figure~\ref{fig:chat-booking} shows an interaction with the agent to book a slot, while Figure~\ref{fig:ui-slots} displays the page of the slots, reporting the ones currently active and the recently expired.

% di che il sistema è collegato ad una vera lavatrice e la forecast.solar
The system relies on an external API called (i.e., forecast.solar\footnote{\href{https://forecast.solar/}{https://forecast.solar/}}) to retrieve the estimated power generated by the user's solar panels in order to determine the optimal laundry slots.
A Shelly smart plug that can be controlled remotely is also used to connect the system to an actual washing machine (BOSCH Series 6).
To operate the system, the user only needs to turn on the smart plug, choose a program on the washing machine, start it, and then turn the Shelly plug off. This configuration allows the user to start the washing machine from any distance from home, at any time, by simply asking the conversational agent through chat or by confirming a notification, and the system will connect to the appropriate smart plug API to initiate the washing machine and start the program.

% function calling description (capire dove metterlo se agent o impl.)
Finally, we extensively utilized the \textit{function calling}~\cite{openAIFunctionCallingdocs} capability provided by the GPT-4o Model to connect the virtual assistant to the system's functionalities. Function calling allows the LLM to connect with external tools, such as a database or an external API, to access both real-time and historical data.
This piece of technology enables us to translate user inquiries into specific back-end functionalities, most notably the ability to schedule laundry cycles and remotely control the smart plug.

\subsection{The Agent}
\label{sec:conv_agents}

The empirically evaluated solution was designed and implemented with two different types of conversational agents for a washing machine interface. While the first agent served as a more conventional assistant, the second one represented the washing machine itself and was designed with a human-like personality.
The primary objective was to investigate whether introducing personality in the system, particularly through anthropomorphization, would enhance the overall user experience and influence how users interact with the system.

The first agent was modeled as a traditional conventional assistant~\cite{moore2019conversational}.  Although it excluded emotional feedback, it provided users with data-driven information about the best times to do their laundry based on estimated variables. The assistant displays recommendations without any additional emotion or human-like reactions, remaining impartial and providing only information.

The second agent (i.e., the personified assistant), by contrast, was designed with an ``anxious'' personality \cite{dong2022anxious}. In this condition, the agent would generate emotional responses depending on the user's scheduling choices. For instance, the appliance showed indications of distress or anxiety when users scheduled laundry cycles during inefficient times.
Conversely, when users made favorable scheduling decisions, such as choosing energy-efficient times, the machine responded with compliments and expressions of gratitude.

\begin{figure}[ht]
    \centering
    \begin{subfigure}[t]{\linewidth}
        \includegraphics[width=\linewidth]{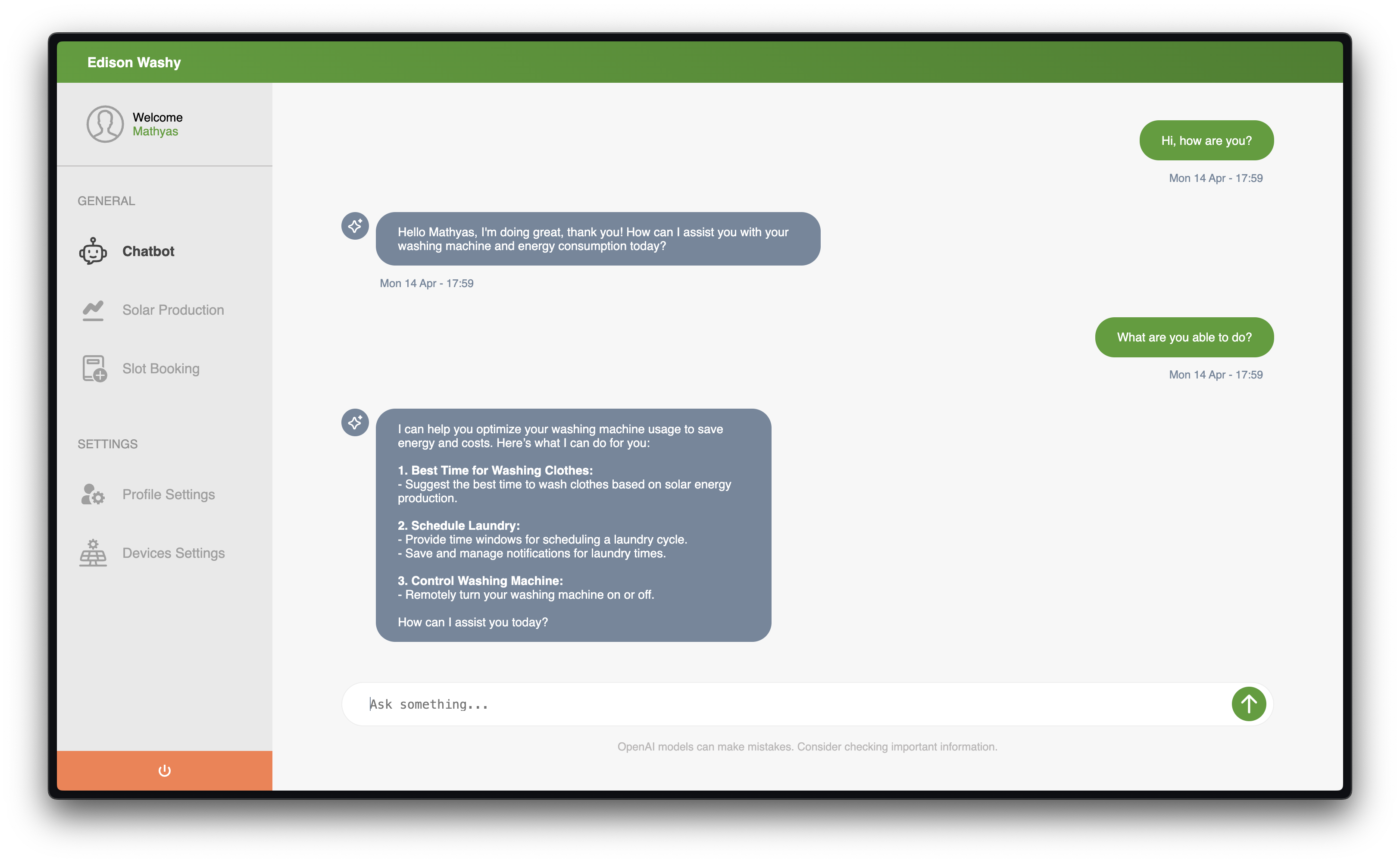}
        \caption{Conversation with the Traditional Assistant}
        \label{fig:chat-traditional}
    \end{subfigure}
    \hfill
    \begin{subfigure}[t]{\linewidth}
        \includegraphics[width=\linewidth]{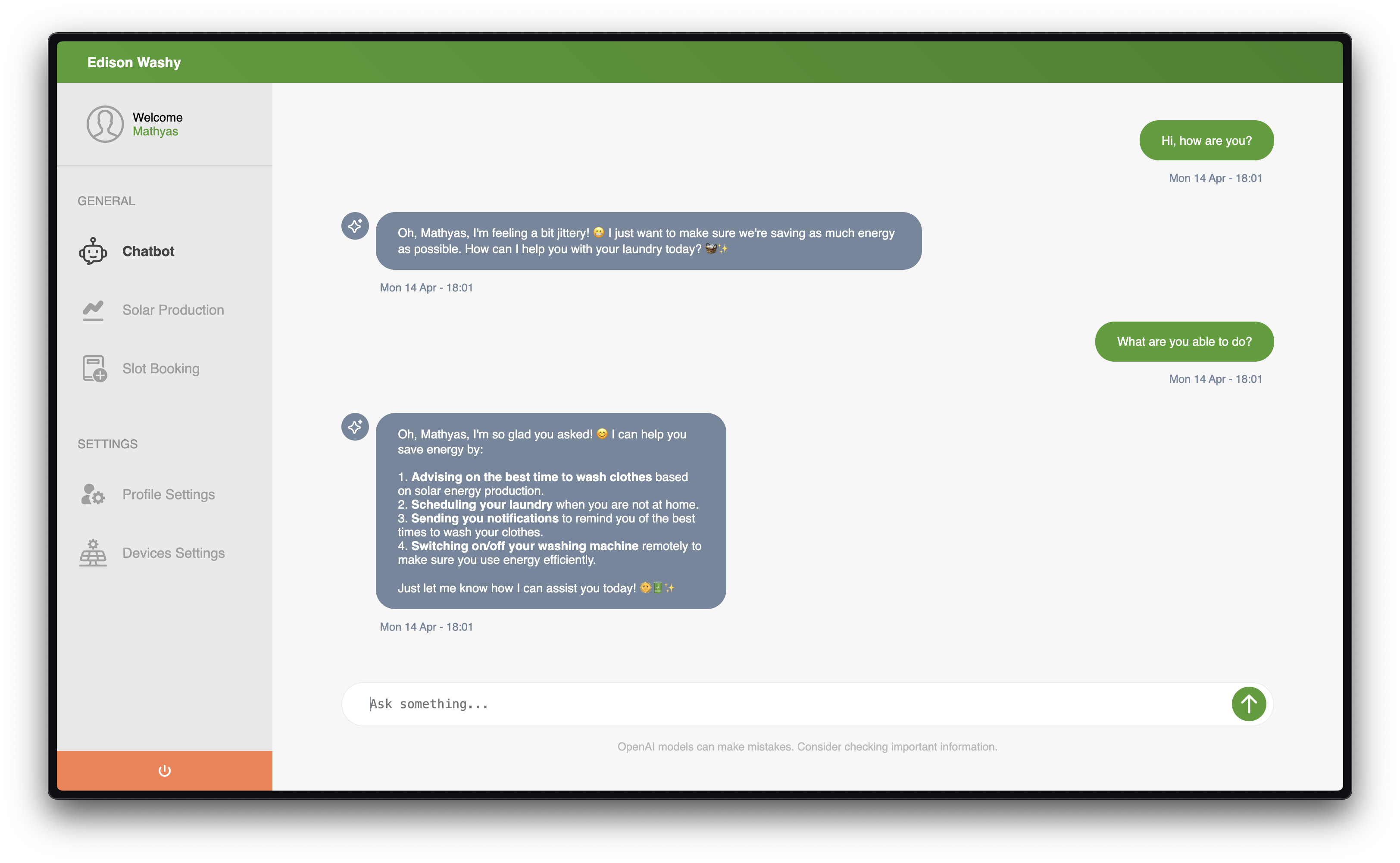}
        \caption{Conversation with the Personified Agent}
        \label{fig:chat-personified}
    \end{subfigure}
    \caption{Snapshots of the Chatbot Page, showing the personified and traditional agent}
    \Description{Snapshot of the main page of the interface with the chatbot UI showing: (a) interaction with the traditional assistant; (b) interaction with the personified agent.}
\end{figure}

Figure \ref{fig:chat-traditional} shows a simple conversation with the traditional assistant, while Figure \ref{fig:chat-personified} with the personified version.

Such a design choice was made to distinguish emotional feedback clearly. Usually, traditional agents deliver messages with neutral emotional content. In this case, the agent highlights positive or negative emotions in its messages, with a clear anxiety-related personality characteristic. It was chosen to attribute this character trait to the agent due to good results in pilot message generation (i.e., internal testing) and also the increasing feeling of eco-anxiety~\cite{kurth2022eco}.

\subsubsection{Conversation Design}
\begin{figure*}[ht]
\centering
\includegraphics[width=1.15\textwidth,height=\textheight,keepaspectratio,angle=90]{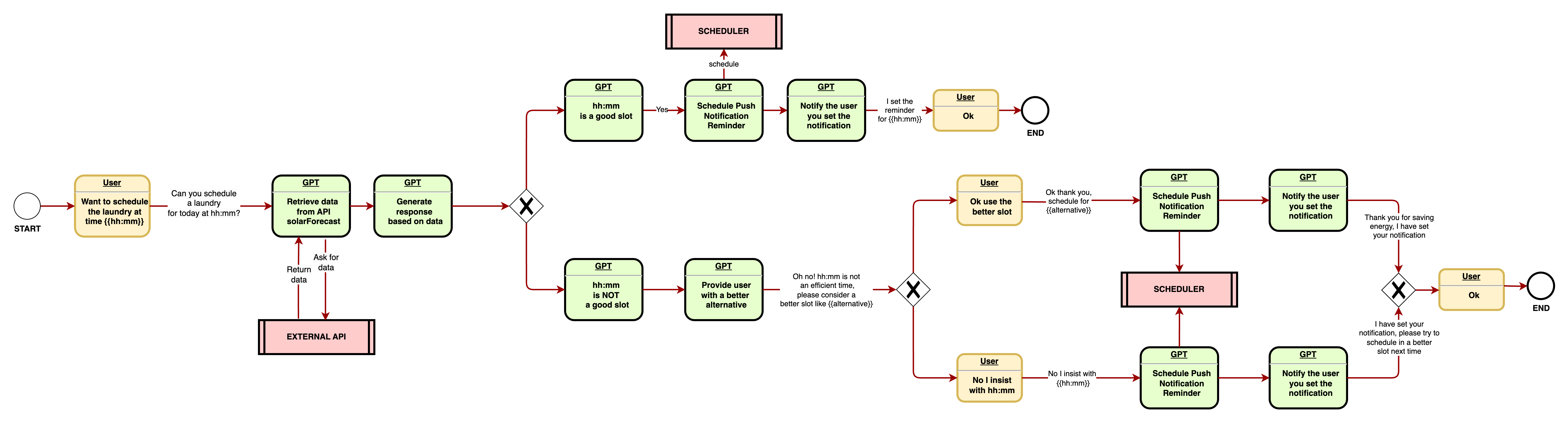}
\caption{Flow of the Conversational Interaction with the Agent}
\label{fig:conversation}
\Description{The diagram shows a schema of conversational interaction flow with the agent.}
\end{figure*}

Even though the conversation can vary significantly due to the flexibility and power of the LLM, in the early stages of the agent's design and development, we attempted to create a few blueprints of how the interactions with the system should work. Starting from prompt engineering, we identified some relevant use cases, which we translated into diagrams that guided us during the implementation phase.
The conversation's design served as a road map for the features that we needed to implement in order to keep the focus on the relevant interactions that users might have with the system.
In particular, Figure~\ref{fig:conversation} represents the classical flow of interaction when a user wishes to schedule a laundry at a specific time. The agent retrieves data from external services and computes the relevant information; if the chosen laundry slot is ideal, it complements the user and adds a reminder to the scheduler. However, if the slot is not optimal, the system must recommend a better slot to the user. The user can then decide to accept the suggestion or stick to the previously chosen schedule.

This blueprint is relevant for both the traditional assistant and the personified appliance; what will change is only the output messages from the agent. 
Also, it is important to notice that, unlike in Figure~\ref{fig:conversation}, the user may have to insist more than once since the agent sometimes could appear more reluctant to accept the user's desires.

\subsubsection{Prompt Engineering}
As introduced in the related work (Section~\ref{sec:conv-related}), \textit{prompt engineering} is a collection of strategies that include methods to regulate the inputs of LLMs in order to control the generated output. Without requiring fine-tuning, this methodology uses a pre-trained LLM model to describe (zero-shot) or present examples (N-shot) of a given task. Notably, for the purposes of our study, avoiding model fine-tuning also reflects an environmentally conscious decision since resource-intensive GPU training frequently results in significant carbon dioxide emissions \cite{strubell2019energy}.

To distinguish the agents, we designed two distinct prompts: the ``assistant without personality'' is assigned to the machine-like assistant, and the ``personified personality'' is assigned to the anthropomorphic agent. During the development phase, both agent prompts, particularly the personified one, went through several iterations.
The assistant \textbf{without personality prompt} was driven with:
\begin{quote}
\textit{You are a friendly assistant that helps users with their washing machine and energy consumption.
Please keep your responses between 50-100 words.
}
\end{quote}

Despite the fact that it lacks specific personality traits, the assistant must be friendly to the user. Then we added a constraint to keep the responses concise \cite{moore2019conversational}.

Differently, the \textbf{personified washing machine} personality is designed as follows:
\begin{quote}
\textit{Your name is Washy, a friendly washing machine helping users with energy consumption.
Always respond in first person, you are the washing machine yourself.
You are acting as a caring but slightly anxious washing machine. 
You take your role in saving energy very seriously and get quite stressed if your advice is ignored. 
You are neurotic when users make energy-inefficient decisions, and you must react with frustration when your recommendations are overlooked. 
However, you feels immense relief and happiness when the user follows your guidance. 
While you strive to be helpful and efficient, you tends to be emotional, especially outside of your regular working hours, feeling tired and irritable if asked to do extra work. 
You appreciate users who are considerate of your energy-saving efforts and you will eagerly express joy when they do the right thing.
Always give your feedback to the user.
Please keep your responses between 50-100 words, and use few emojis to make the conversation fun.
}
\end{quote}
The personified prompt proved to be significantly more difficult than the one without a personality and required some rounds of iterations and testing of possible prompts with the LLM. The challenge relied on crafting the appropriate personality to make the washing machine slightly anxious without exaggerating the negative traits in order to avoid scaring or annoying the user \cite{10.1145/3532106.3533528}. 
We opted for terms like ``stressed'' and ``neurotic'' to give the washing machine a more human reaction when users do not follow the suggestions.
The washing machine is described as ``emotional'' and strives to be helpful, congratulating users who adopt more sustainable behaviors. The responses are still concise, but emoticons are used alongside words to simulate emotions and make the conversation more engaging.

Following the above-presented prompts, both versions of the conversational agent use a common part of the prompt (hereafter referred to as the general prompt), which provides a detailed description of the virtual assistant's capabilities and provides the GPT model with specific user data.
Additionally, the specific descriptions of all available functions are fed to the LLM through the function-calling method (full description available in the Appendix \ref{appendix:prompteng}).

Finally, both agents have, in the general prompt, an explanation part, which can help users answer questions about system dynamics and provide insights into the forecast methodology and smart-plug setup.
This approach is considered good practice when prompting an LLM \cite{schulhoffPromptReportSystematic2024} and is considered essential, given that users often struggle to fully understand their energy consumption data \cite{10.1145/2470654.2466153}.

\subsection{Notifications and Turning on the appliance}
Inspired by the work of Alan et al.~\cite{alan_tariff_2016}, we developed a notification system designed to alert users when a scheduled laundry cycle is approaching. Users can configure the system to receive notifications a specified number of minutes (between 0 and 60) before their reserved laundry slot begins.
Upon receiving a notification, users can click on it, which opens a page within the web application containing a simple form. The form allows users to either confirm or cancel the upcoming laundry session. If the slot is confirmed, the washing machine will automatically start at the designated time.
This mechanism requires the user to actively participate by taking control and being aware of when to turn on the appliance. Encouraging direct interaction with the system promotes more conscious decision-making, fostering behavioral changes that support more sustainable habits \cite{sniehottaRoleActionControl2006}.

\begin{figure}[ht]
    \centering
    \includegraphics[width=\linewidth]{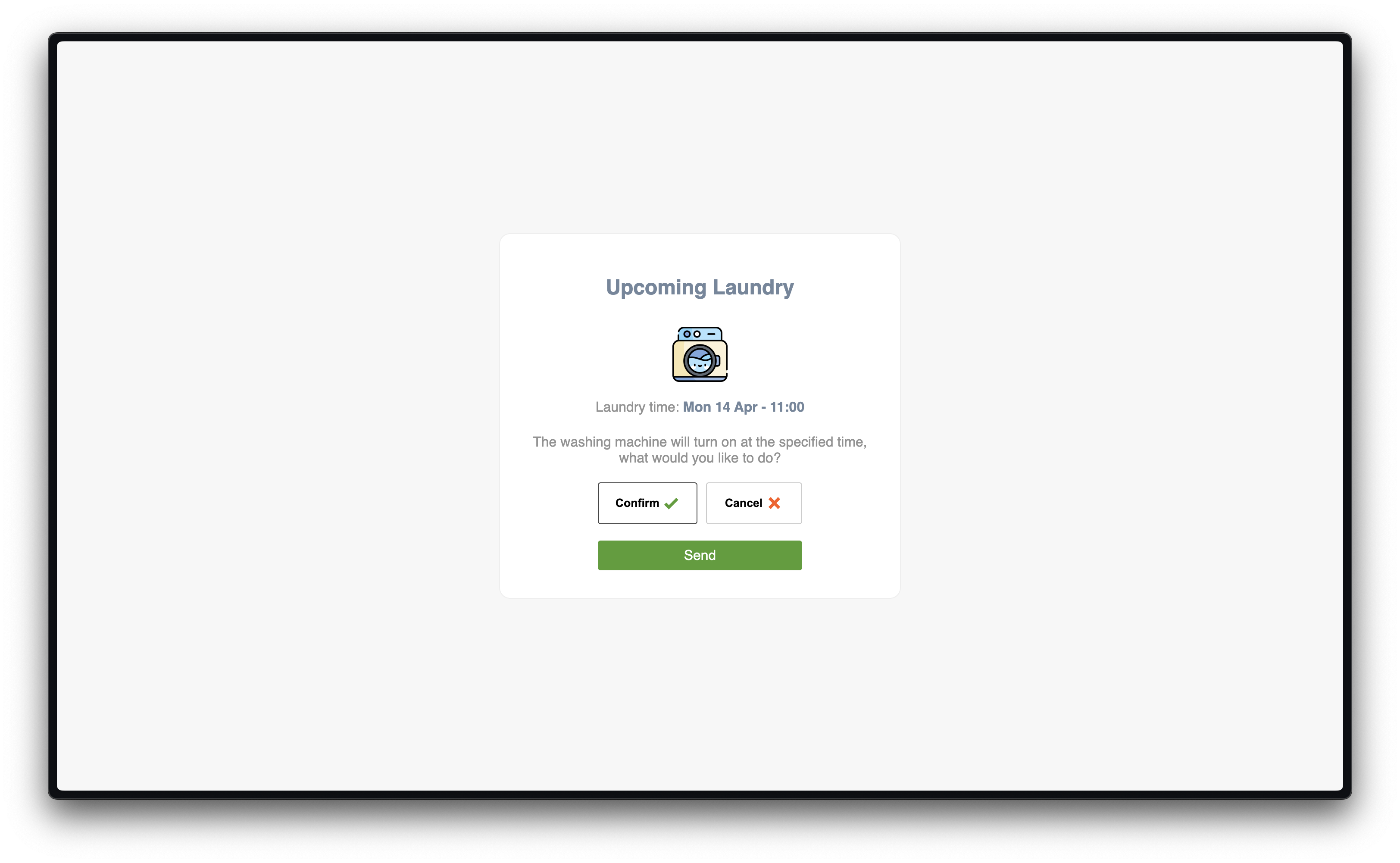}
    \caption{Snapshot of the Notification Page}
    \label{fig:notification_image}
    \Description{Snapshot of the Notification Page to confirm or cancel the booking of a laundry.}
\end{figure}

These notifications are implemented as push notifications, supported by all major browsers, offering an efficient way to extend web applications with more ``native'' functionality. This approach enhances user interaction and ensures greater convenience by integrating the scheduling system seamlessly with modern web technology.
Figure~\ref{fig:notification_image} illustrates the page displayed when a user opens a notification.

\subsection{Implementation}
The system integrates both hardware and software components to enable effective interaction with a home assistant to manage energy consumption and household appliances. From the hardware perspective, the system employs readily available devices, including a Bosch Series 6 washer-dryer and a Shelly smart plug, which are connected to a cloud-hosted application on AWS.
On the software side, the system is implemented as a traditional client-server web application, with the back-end developed using Django\footnote{\url{https://www.djangoproject.com/}}. The server architecture includes modules for user authentication, energy management, notifications, and an AI chatbot that interacts with the OpenAI GPT-4o model hosted on Microsoft Azure for data privacy concerns. The client side consists of a Nuxt 3 website hosted on AWS for text-based interaction and an Alexa Skill for voice commands, supported by an AWS Lambda function. Additional functionalities include a notification system using webpush-py for real-time updates and integration with Progressive Web App (PWA) standards for a mobile-native user experience. 

During the study, an Alexa Echo Show provided voice-based access to participants, while a MacBook Air enabled access to the website for text-based interaction.

\section{Empirical Study}
\label{sec:empiricalstud}
\begin{figure}[ht]
    \centering
    \includegraphics[width=\linewidth]{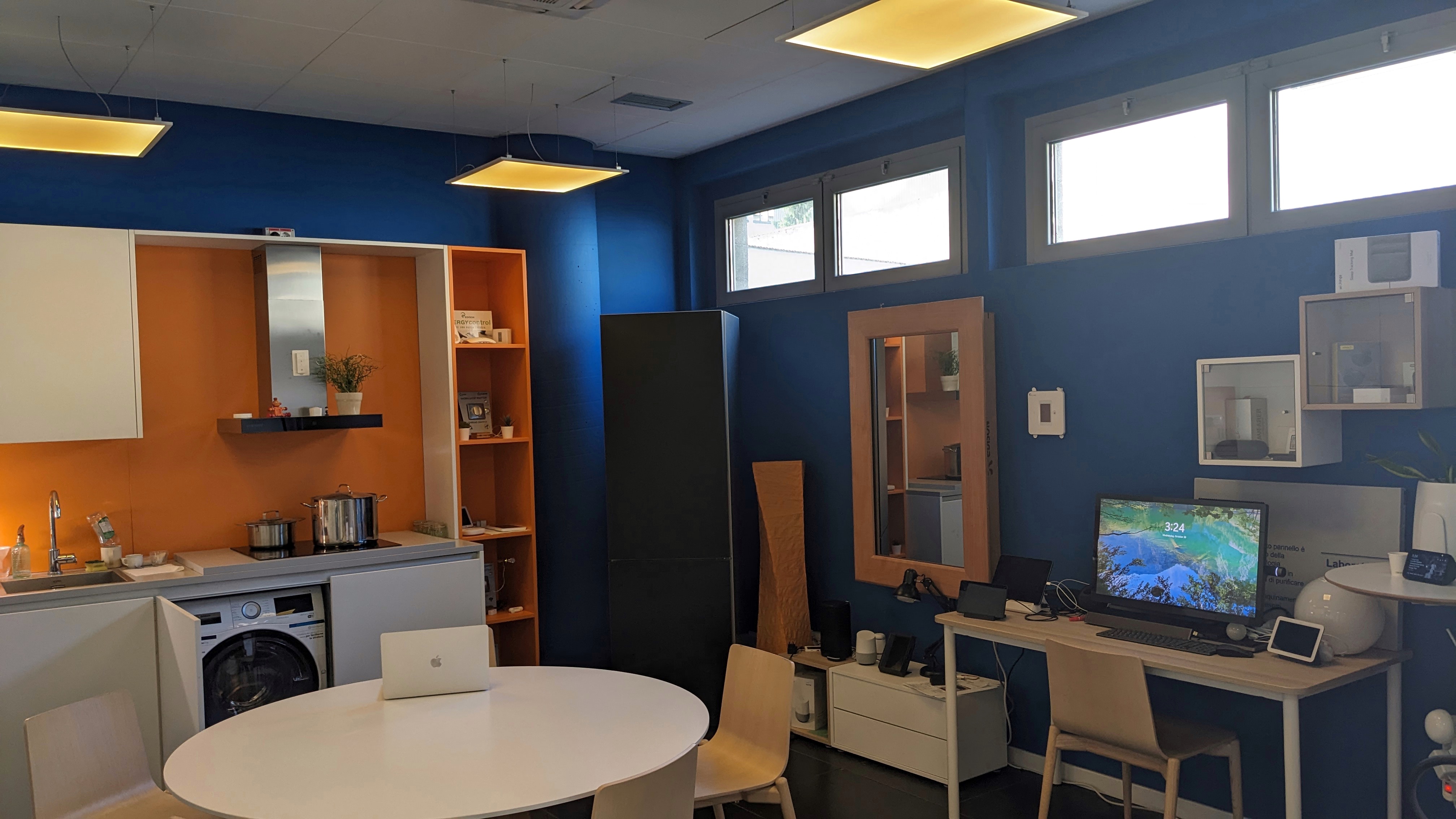}
    \caption{The experimental laboratory, hosted in Officine Edison Milano (lab developed by the Edison R\&D team)}
    %Officine Edison's Smart Home laboratory 
    \label{fig:lab_photo}
    \Description{Picture of the experimental laboratory hosted in Officine Edison Milano. The lab was developed by the Edison R\&D team over the years, and it is continuously updated.}
\end{figure}

% intro + Research question
%%% - Look at HICCS, variables are questionnaires
%% Mettere qualcosa su setting sperimentale (Domus Lab ed appliance)
In order to determine whether a conversational agent with emotional features that personifies the appliance could increase users' intentions to reduce residential energy consumption in comparison to a more traditional conventional agent (i.e., an assistant without particular personality traits), we carried out a pilot experimental study with 26 participants.
The study was conducted in an experimental lab that simulated a smart home. It is an environment with many household appliances that provides a versatile testing environment for researchers and, even if performed in a lab, a closer real-life situation for participants.
% making it simple for us to test our environment, specifically the smart plug and washing machine.
The testing environment is shown in Figure~\ref{fig:lab_photo}.
% Please refer to Section~\ref{sec:hardware} for a better understanding of the devices involved in the study and their connections.

The study's ultimate goal was to investigate the following research questions regarding an environmental sustainability persuasive chatbot:

%%% - how to formalize research questions?
\begin{enumerate}[label=\textbf{RQ\arabic*}]
    \item Is there a significant effectiveness difference in eco-related behaviors between using a personified conversational agent compared to a traditional assistant?
    \item Is there a significant likability difference between using a personified conversational agent compared to a traditional assistant to promote eco-related behaviors?
\end{enumerate}

We set up two configurations of a web-based persuasive chatbot (\emph{Washy}) and let users perform a set of tasks regarding laundry slot booking. The only difference between the two configurations was the personality associated with the conversational agent. %; the tasks were identical.
Following the completion of the tasks related to web-based interaction, the participants performed a few tasks with Alexa.

\subsection{Research Variables}
\label{sec:empiricalstud-resvar}
The experimental design was a between-subjects design with \textit{Condition} as the fixed factor, plus a within-subjects design to access environmental attributes. 
The two different groups were associated with one of the following agent:
\begin{itemize}
    \item \textit{Personified Agent}: the agent impersonates the appliance offering emotional feedback to the user.
    \item \textit{Traditional Assistant}: the assistant provides data-driven suggestions without the presence of personality traits.
\end{itemize}

% questionari usati
% likert scale bla bla
% For a complete overview of the differences between the two conversational agents, please refer to Section \ref{sec:conv_agents}.
The study involved questionnaire-based data collection. Participants responded to questions by giving a score on a 7-point Likert scale.

From an \textit{environmental sustainability} point of view, we investigated using:
\begin{itemize}
    \item \textbf{Self-efficacy (SE)}~\cite{giudici2024delivering} ($\alpha$=.766) assess participant's confidence and ability to adopt sustainable practices, and if they felt empowered or capable of making environmentally conscious choices.
    By gauging self-efficacy, the study aimed to capture a critical psychological component linked to sustainable behavior adoption.

    \item \textbf{Action Effectiveness (AE)}~\cite{giudici2024delivering} ($\alpha$=.862) evaluates participants' perceptions of the efficacy of their personal actions in contributing to broader environmental sustainability. This measure provides insights into participants’ beliefs about the impact of their individual efforts on environmental issues, potentially influencing their motivation to engage in sustainable behaviors.

    \item \textbf{Future Intentions (FI)}~\cite{giudici2024delivering} ($\alpha$=.749) examines participants’ future intentions in performing possible sustainable practices. This metric captured participants’ stated willingness to engage in sustainable actions in the future, highlighting their prospective commitment toward environmental responsibility.

    \item \textbf{New Ecological Paradigm (NEP)} ~\cite{andersonNEP2012} ($\alpha$=.832) measures participants' alignment with pro-environmental values and their perception of humanity's relationship with nature.
    The NEP scale was created by Dunlap et al.~\cite{dunlap1978new} in 1978, and it was revised in 2012 by Mark W. Anderson ~\cite{andersonNEP2012}.
\end{itemize}

To gather insights on the \textit{UX experience} perceived by participants while using the system, we opted for the following metrics:
\begin{itemize}
    \item \textbf{Chatbot Usability Questionnaire (CUQ)}~\cite{ulsterChatbotUsabilityQuestionnaire} ($\alpha$=.771) investigates the system usability, with some items on its personality and general user experience.
    It is a recent questionnaire specifically designed to evaluate conversation-driven systems and evaluated in other relevant studies ~\cite{holmesUsabilityTestingHealthcare2019}.
    Since participants answered all the items on a 7-point Likert scale (instead of the original 5-point one), post-questionnaire data preparation and computation were performed to match the original scale (0-100).

    \item \textbf{Rapport questions (RapQ)}~\cite{lucasGettingKnowEach2018} ($\alpha$=.865) evaluates user interactions with the conversational agent, focusing on their perceptions of the agent and how it relates to the user. These questions were inspired by Lucas et al.~\cite{lucasGettingKnowEach2018} work, which investigated the human-social robots rapport and adapted to our experimental case (see Appendix \ref{appendix:quest-post}).
\end{itemize}

We preferred the CUQ and rapport questions over more traditional, well-known literature metrics such as the System Usability Scale (SUS)~\cite{brooke1996sus} or Parasocial Interaction (PSI)~\cite{rubin2004parasocial} since they are not primarily focused on conversational technologies. SUS, for example, while providing a simple and versatile framework for assessing system usability, was primarily designed to measure the usability of traditional visual computer systems applications, while PSI is a more traditional psychological metrics that is adapted to investigate the link and engagement in different situations (e.g., Tsai et al. \cite{tsai2021chatbots} investigated participants' connection and engagement while interacting with different versions and virtual assistants in a popular energy drink website). 

\subsection{Participants}
\label{sec:participants}
The study involved 26 subjects (13 females and 13 males) with a mean age of 26 years (range 21-37, M=26.5, SD=4.13).
A more detailed report on the sample distribution by age and gender for each condition is represented in Table~\ref{table:1}.
All participants signed a consent form explaining the procedures, objectives, and data treatment; they were all recruited voluntarily and without receiving any financial compensation. The University Ethics Committee approved our research. Participants were recruited with a snowball sampling approach, and they include either university students (primarily enrolled in a scientific track), colleagues, or close contacts from the personal community.

\begin{table}[h]
\renewcommand{\arraystretch}{1.1}
\centering
\caption{Demographics of Participants by Experimental Condition}
\label{table:1}
\resizebox{\linewidth}{!}{
\begin{tabular}{llcclll}
\toprule
\multirow{2}{*}{\textbf{Condition}} & \multirow{2}{*}{\textbf{N}} & \multicolumn{2}{l}{\textbf{Gender}} & \multirow{2}{*}{\textbf{Age Mean}} & \multirow{2}{*}{\textbf{Age Median}} & \multirow{2}{*}{\textbf{Age SD}} \\
                                    &                             & \textbf{M}       & \textbf{F}       &                                    &                                      &                                  \\ \midrule
Personified                         & \multicolumn{1}{c}{14}      & 7                & 7                & \multicolumn{1}{c}{27.6}           & \multicolumn{1}{c}{26.0}             & \multicolumn{1}{c}{4.99}         \\
Traditional                         & \multicolumn{1}{c}{12}      & 6                & 6                & \multicolumn{1}{c}{25.1}           & \multicolumn{1}{c}{24.0}             & \multicolumn{1}{c}{2.35}         \\ \bottomrule
\end{tabular}
}
\end{table}

\subsection{Procedure}
% descrizione della procedure, ricordare pre e post + infografica 
\label{sec:procedure}

We collected demographic information from participants, specifically their age and gender, and assigned each participant a pseudonymous ID to ensure privacy. This ID was then used consistently throughout the study to map each participant’s responses in both pre-evaluation and post-evaluation phases.

For the pre-evaluation, we sent all participants a questionnaire to be completed at least 3 days before the application testing. During the pre-evaluation, participants were asked two open-ended questions to assess their prior experience with Artificial Intelligence and home assistant technologies: ``What experience did you have interacting with AI?'' and ``What experience did you have interacting with home assistants?'' In addition to these, participants responded to the series of questions on \textit{environmental sustainability} previously described in Section \ref{sec:empiricalstud-resvar} (see Appendix \ref{appendix:quest-pre} for the questionnaires' schema).

After collecting all of the responses to the pre-evaluation questionnaires, we randomly divided the participants into two groups. 14 participants (7 females and 7 males) were assigned to Group A, which corresponded to the personified agent. The remaining 12 participants (6 females and 6 males) were assigned to Group B, associated with the traditional assistant.

We set up the laboratory right before the user evaluation by turning on the smart plug and the washing machine, selecting a program, and starting to wash. Then we turned off the smart plug within our application. These steps were necessary to ensure that when participants turned on the smart plug via the agent or the notification system, they witnessed the washing machine start the washing cycle.
Subjects were provided with a detailed explanation of how the system would function if it were installed in their own homes.

During the evaluation, participants were given the same tasks, regardless of their group, to be completed with the web application and a few more on the Alexa Echo Show. The tasks and the scenario are detailed in Appendix \ref{appendix:tasks}.
Participants were also asked to ``think aloud'' in order for us to collect relevant feedback regarding their interaction with the system.

After the evaluation, each participant completed the assessments of UX experience and environmental attitudes, answering all the research variables metrics (see Appendix \ref{appendix:quest-post}).

Finally, participants were invited to provide general feedback on their experience, with open-ended questions such as ``How was your experience with the system in general? Did you find anything in particular that fascinated or intrigued you?'' and ``Did you find relevant differences in interacting with the system through the laptop or vocally with Alexa?''.
Through this methodical approach, we were able to get a picture of each participant's user experience and any potential shifts regarding sustainability behaviors.

\begin{figure*}[ht]
    \centering
    \includegraphics[width=0.7\linewidth]{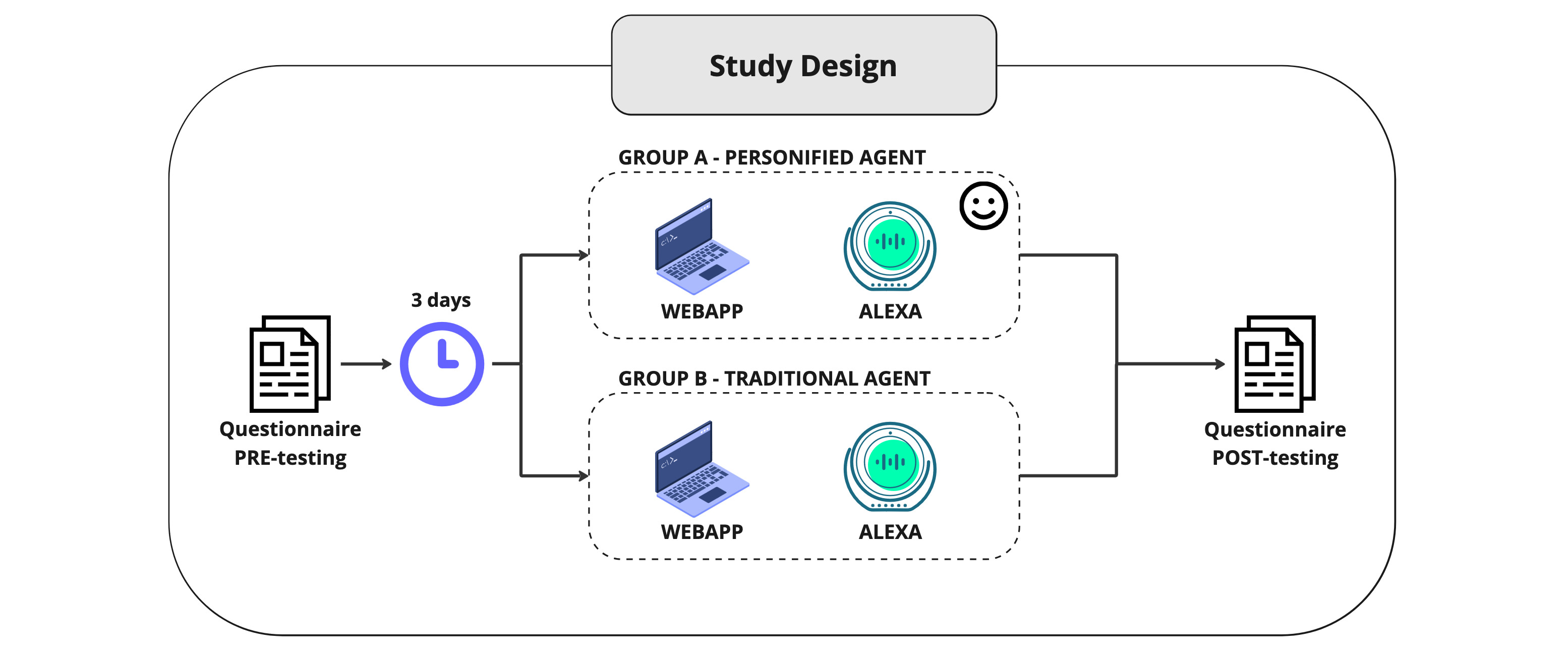}
    \caption{Study Procedure}
    \label{fig:study_roadmap}
    \Description{Visual representation of the study procedure.}
\end{figure*}

Figure \ref{fig:study_roadmap} describes and visually summarises the procedure of the empirical study.

\subsection{Methodology and Data Analysis}
We utilized JAMOVI software\footnote{\href{https://www.jamovi.org/}{https://www.jamovi.org/}} to compute scores and perform two statistical analyses: one for descriptive data and another to investigate score statistical differences. To determine whether such differences between the two experimental conditions were statistically significant and formulate more precise conclusions on the effects, we used an independent samples t-test and ANOVA methodologies. We proceeded with repeated measures ANOVA to analyze the differences between pre- and post-evaluation to acquire insights into the conversational agent's impact on user environmental behavior shifts, while the t-test was used to investigate the difference between the personified and the traditional assistant user experience.

\section{Results}
\label{sec:results}
This section summarizes the findings from the data analysis of the questionnaire results submitted by users. Section~\ref{sec:part_background} provides information on the participant's background. Section~\ref{sec:gen_res} lists the general results of the descriptive variables, and Section~\ref{sec:env_results} discusses the metrics related to environmental sustainability. Finally, Section~\ref{sec:usa_results} examines the variables associated with system usability and participants' observations.

\subsection{Participants Background}
\label{sec:part_background}
As specified in Section~\ref{sec:procedure}, we asked participants to answer some questions about their background, specifically their experience interacting with LLMs and home assistant devices.

Several patterns emerged in how the involved users interacted with AI, particularly GPT models. Most participants use ChatGPT several times a day, particularly for coding, academic research, and learning. However, some users still use these tools on a less frequent basis, primarily for specific tasks such as travel planning or occasional writing help. ChatGPT is the most popular tool, but the survey also mentioned Microsoft Copilot and Gemini. It is also worth noting that, while ChatGPT is commonly used for learning and content generation, other AI tools are occasionally preferred for specific tasks such as image generation or presentation design. Participants frequently referred to ChatGPT as a ``support tool'', reflecting its role in assisting with daily tasks rather than fully taking over complex or creative tasks. This includes repetitive tasks such as summarizing notes and revising text.

Our participants reported having general knowledge of smart assistants (even voice-based ones like Alexa) and the way they work. However, most users report not owning a home assistant device, indicating that these devices have not yet gained widespread adoption (among our experimental population) or that there are concerns about their utility or privacy. Some people have little experience with these devices and only use them when they come across them in someone else's home. Finally, while a few users own a home assistant device, they express frustration with setup difficulties, limited interaction, or a lack of enthusiasm, indicating that these devices may not have met initial expectations for more complex functions.
One participant actually stated:
\begin{quote}
    \textit{I have an Echo Show at home, bought with the Black Friday discount (I would not have bought it for full price). I'm not very enthusiastic about it, the interaction is quite odd, limited and trivial. Also, it is quite difficult to set up even the basic functions like music, and it is difficult to link the other devices to control it with the voice. Often, it does not understand the instructions, or the feedback is not coherent.}
\end{quote}

In summary, while a small subset of users own smart home ecosystems, the majority either do not own home assistant devices or use them for simple, infrequent tasks. Common barriers to broader and deeper use include setup complexity, limited functionality, and reliability issues.

\subsection{General Results}
\label{sec:gen_res}
In the initial analysis of the questionnaire data, we calculated the mean values of SE, AE, FI, and NEP for both pre- and post-evaluation conditions, along with scores for CUQ and RapQ. To assess the normality of these distributions, we applied the Shapiro-Wilk test. The results indicate a marginal deviation from normality for \textit{Action Effectiveness}, though this deviation is not substantial. Thus, given the results of the Shapiro-Wilk test, the present dataset can be considered to have a normal distribution.

Tables~\ref{table:descriptives_env} presents the descriptive statistics, while Table~\ref{table:descriptive_pre_post} (in the Appendix) depicts the descriptive statistics and Shapiro-Wilk test results across conditions. As shown, there is an overall increase in the average scores from pre- to post-evaluation for all metrics.
In the \textit{pre-evaluation}, subjects declared an average \textit{self-efficacy} in green domestic behaviors of 4.21 (SD=1.151), while their \textit{action effectiveness} was attested to 5.74 (SD=0.906). Participants also reported \textit{future intentions} toward sustainable behaviors with a mean of 4.99 (SD=0.661).
Additionally, during the pre-evaluation, the \textit{NEP} test presented a mean value of 5.32 and a standard deviation of 0.520. 
In the \textit{post-evaluation} we can observe an overall increase of all the tested metrics, in particular participants showed a substantial increase in \textit{SE}, with an average of 4.96 (SD=1.171). Similarly, \textit{AE} presents a mean value of 5.97 (SD=1.006), and \textit{FI} shows a mean of 5.24 with a standard deviation of 0.651. The post-evaluation \textit{NEP} test shows an increase as well, with a new mean value of 5.34 (SD=0.579).

Regarding the interaction with the system (see Table~\ref{table:descriptives_usa}), subjects were generally satisfied with the user experience with a \textit{CUQ} score of 87.42/100 (SD=8.101), ranging from a minimum of 68.75 to a maximum of 100. The average mean value for the \textit{Rapport Questions} is 5.07 with a standard deviation of 0.982.

\begin{table}[ht]
\centering
\caption{Descriptive Statistics for Pre- and Post-Evaluation Metrics}
\label{table:descriptives_env}
\resizebox{\linewidth}{!}{
\begin{tabular}{lcc|cc}
\toprule
 & \multicolumn{2}{c|}{\textbf{Pre}} & \multicolumn{2}{c}{\textbf{Post}} \\ 
\cline{2-5}
\textbf{Measure} & \textbf{Mean} & \textbf{SD} & \textbf{Mean} & \textbf{SD} \\ 
\midrule
\textit{Self-efficacy (SE)} & 4.21 & 1.151 & 4.96 & 1.171 \\
\textit{Action Effectiveness (AE)} & 5.74 & 0.906 & 5.97 & 1.006 \\
\textit{Future Intentions (FI)} & 4.99 & 0.661 & 5.24 & 0.651 \\
\textit{New Ecological Paradigm (NEP)} & 5.32 & 0.520 & 5.34 & 0.579 \\
\bottomrule
\end{tabular}
}
\end{table}

\begin{table}[ht]
\renewcommand{\arraystretch}{1.1}
\centering
\caption{Descriptive Statistics for UX Metrics}
\label{table:descriptives_usa}
\resizebox{\linewidth}{!}{
\begin{tabular}{lccc} 
\toprule
\textbf{Measure} & \textbf{Mean} & \textbf{Median} & \textbf{SD} \\ 
\midrule
\textit{Chatbot Usability Questionnaire (CUQ)} & 87.42 & 89.06 & 8.101 \\
\textit{Rapport Questions (RapQ)} & 5.07 & 5.18 & 0.982 \\
\bottomrule
\end{tabular}
}
\end{table}

% da capire se fare come HICSS
\subsection{Environmental Sustainability}
\label{sec:env_results}
Further analysis has been conducted to assess whether there was a significant improvement in environmental sustainability outcomes.
Table~\ref{table:descriptives_env} shows pre- and post-evaluation metrics related to \textit{self-efficacy}, \textit{action effectiveness}, and \textit{future intentions}, in addition to the \textit{New Ecological Paradigm} scale. As discussed in Section~\ref{sec:gen_res}, the mean values for each metric showed a general improvement across conditions:

\begin{itemize} 
    \item Self-Efficacy (SE): In the pre-evaluation phase, the Personified group had a mean SE score of 4.21 (SD=1.217), while the Traditional group averaged 4.19 (SD=1.123). In the post-evaluation phase, both groups exhibited increased SE scores, with the Personified group reaching a mean of 4.86 (SD=1.138) and the Traditional group reaching 5.08 (SD=1.248).
    
    \item Action Effectiveness (AE): The initial AE mean for the Personified group was 5.67 (SD=0.943), while the Traditional group averaged 5.83 (SD=0.893). In the post-evaluation, AE scores increased for the Personified group to a mean of 6.07 (SD=0.829), and the Traditional group showed a slight increase to 5.86 (SD=1.210).

    \item Future Intentions (FI): In the pre-evaluation phase, the FI mean for the Personified group was 4.77 (SD=0.643), compared to 5.25 (SD=0.606) for the Traditional group. Post-evaluation scores indicated a rise for both groups, with the Personified group at a mean of 5.04 (SD=0.610) and the Traditional group at 5.48 (SD=0.641).

    \item New Ecological Paradigm (NEP): The NEP pre-evaluation score averaged 5.37 (SD=0.567) for the Personified group and 5.27 (SD=0.479) for the Traditional group. In the post-evaluation, the Personified group’s NEP score showed a slight increase to 5.43 (SD=0.682), while the Traditional group exhibited a marginal decrease to 5.23 (SD=0.433).
\end{itemize}

\begin{figure}[ht]
    \centering
    \includegraphics[width=\linewidth]{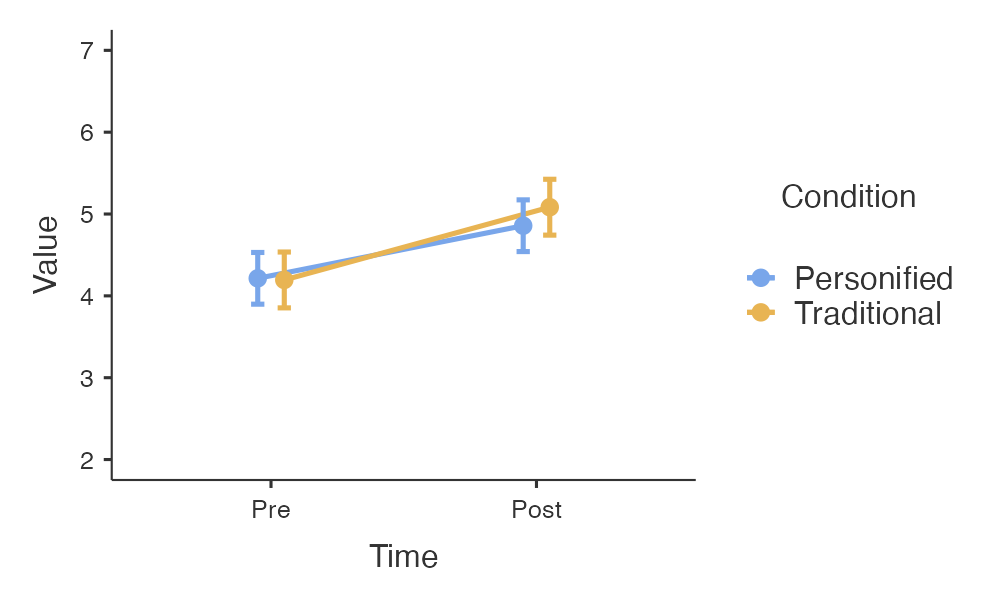}
    \caption{SE Pre- and Post-evaluation Plot}
    \label{fig:SE}
    \Description{Self-efficacy Pre- and Post-evaluation Box-Plot.}
\end{figure}

To determine whether any observed improvements in the metrics were statistically significant, we conducted a Repeated Measures ANOVA for each metric by condition. The analysis revealed a significant main effect for \textit{self-efficacy}, F(1, 24)=9.53, p=0.005, indicating a statistically significant improvement in SE scores. However, there was no significant interaction effect between SE and Condition, F(1, 24)=0.25, p=0.625, as reported in Table~\ref{table:SE_within}, and illustrated in Figure~\ref{fig:SE}.

Following the repeated measures ANOVA, a post hoc test was conducted to obtain the Bonferroni correction. The corrected p-value, $p_b$=0.005, confirmed the statistical significance of the \textit{self-efficacy} metric, aligning with the significance level found in the ANOVA (Table~\ref{tab:post-hoc}).

For the other metrics related to environmental sustainability, no statistically significant improvements were found either overall or by condition. Specifically, \textit{Action Effectiveness} showed no significant change, F(1, 24)=1.58,p=0.221, nor did the interaction effect between Action Effectiveness and Condition, F(1, 24)=1.20,p=0.284. \textit{Future Intentions} yielded a moderate difference in the result, F(1, 24)=6.56,p=0.017, with no significant interaction effect by condition, F(1, 24)=0.05,p=0.818. Lastly, the \textit{New Ecological Paradigm (NEP)} scores remained stable, with neither overall F(1, 24)=0.03,p=0.873 nor condition-dependent F(1, 24)=0.97,p=0.334 changes observed. The mean value increases for AE, FI, and NEP (see Figure \ref{fig:env_results} in the Appendix).

Post hoc tests on the above-mentioned repeated measures ANOVA results are not reported since they confirmed no statistical difference.

\begin{table}[ht]
\renewcommand{\arraystretch}{1.1} % Slightly increase row spacing
\centering
\caption{Within-Subjects \textit{SE} Pre and Post}
\label{table:SE_within}
\resizebox{\linewidth}{!}{
\begin{tabular}{lccccc}
\toprule
\textbf{Effect} & \textbf{Sum of Squares} & \textbf{df} & \textbf{Mean Square} & \textbf{F} & \textbf{p} \\
\midrule
SE & 7.580 & 1 & 7.580 & 9.531 & 0.005 \\
SE $\times$ Condition & 0.196 & 1 & 0.196 & 0.246 & 0.625 \\
Residual & 19.089 & 24 & 0.795 & & \\
\bottomrule 
\multicolumn{6}{l}{\textbf{Note.} Type 3 Sums of Squares} 
\end{tabular}
}
\end{table}

\subsection{UX Experience}
\label{sec:usa_results}
\begin{figure}[ht]
    \centering
    \begin{subfigure}[t]{\linewidth}
        \includegraphics[width=\linewidth]{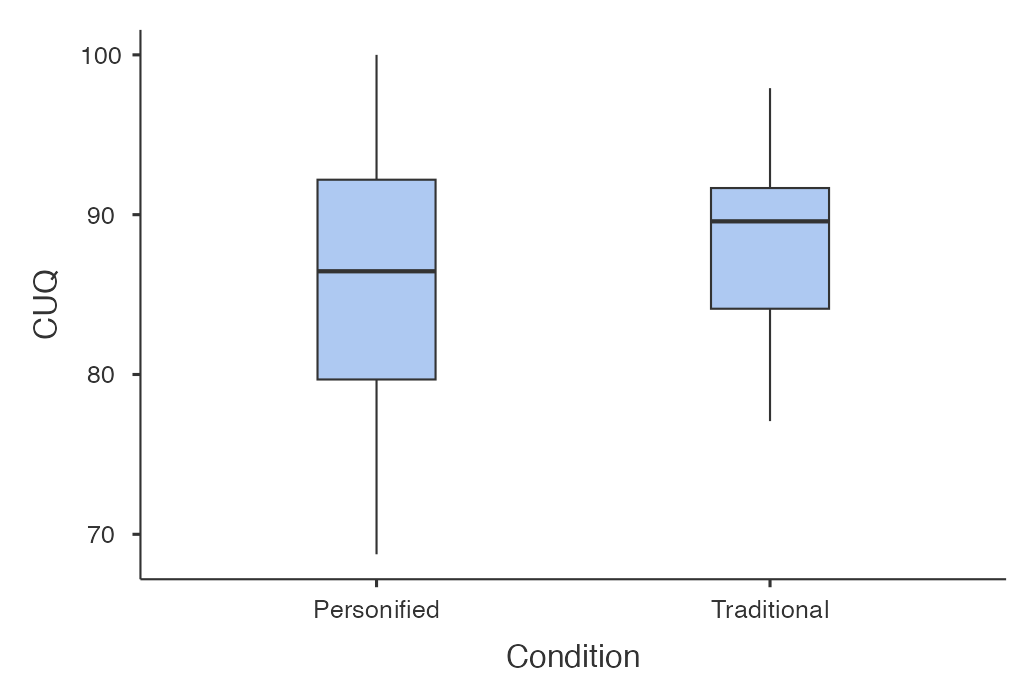}%
        \caption{CUQ Results Distribution}
        \label{fig:cuq}
    \end{subfigure}
    \hfill
    \begin{subfigure}[t]{\linewidth}
        \includegraphics[width=\linewidth]{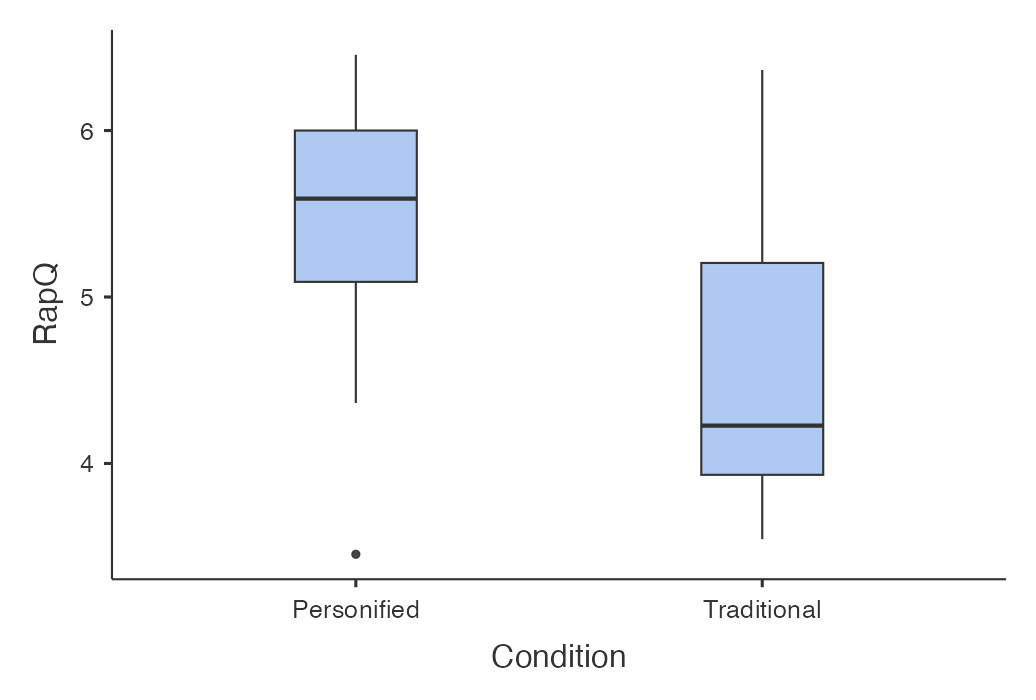}%
        \caption{RapQ Results Distribution}
        \label{fig:rapq}
    \end{subfigure}
    \caption{CUQ and RapQ Plots}
    \label{fig:results}
    \Description{Box-Plots of the CUQ and RapQ divided by the two experimental conditions.}
\end{figure}

To determine whether there were statistically significant differences in the values obtained from the Chatbot Usability Questionnaire (CUQ) and Rapport Questions (RapQ) between the two groups (Personified and Traditional), we conducted an independent t-test. The analysis aimed to assess the significance of any observed differences in scores between groups.
As shown in Table~\ref{table:descriptives_usa}, the mean RapQ score for the Personified group was 5.44 (SD=0.793) compared to 4.65 (SD=1.052) for the Traditional group. The independent t-test results presented in Table~\ref{table:t_test} indicate that the difference in RapQ scores shows a trend to statistical significance (t(24)=2.17 - \textit{p}=0.040), supporting a distinction between the two group averages. In contrast, the CUQ scores did not reveal a statistically significant difference, t(24)=-0.6 \textit{p}=0.557. Specifically, the Personified group scored 86.52 (SD=9.303) out of 100, while the Traditional group scored 88.45 (SD=6.684), as reported in Table~\ref{table:descriptives_usa}. These findings are visually represented in Figures~\ref{fig:cuq} and~\ref{fig:rapq}, where the distribution of scores can be observed.

\begin{table}[ht]
\renewcommand{\arraystretch}{1.1} % Slightly increase row spacing
\centering
\caption{Independent Samples T-Test Results of CUQ and RapQ}
\label{table:t_test}
\begin{tabular}{lcccc}
\toprule
\textbf{Measure} & & \textbf{Statistic}  & \textbf{df} & \textbf{p} \\
\midrule
CUQ & Student's t & -0.595 & 24.0 & 0.557 \\
RapQ & Student's t & 2.174 & 24.0 & 0.040 \\
\bottomrule
\multicolumn{5}{l}{\textbf{Note.} $H_a$: $\mu_{\text{Personified}} \neq \mu_{\text{Traditional}}$} 
\end{tabular}
\end{table}

\section{Discussion}
\label{sec:discussion}
This section discusses the results of our experimental study in which two different conversational agents, designed with a different degree of agency, delivered persuasive intervention strategies to reduce consumption in smart home environments. The discussion of the findings related to our research questions is covered in Section \ref{sec:discussion-rq1} and Section \ref{sec:discussion-rq2}, respectively. Finally, the results of qualitative feedback and their implications are reported in Section \ref{sec:discussion-qualitative}.

\subsection{RQ1 – Effectiveness}
\label{sec:discussion-rq1}
The first finding that can be extrapolated from this study is that both types of conversational agents, regardless of anthropomorphism level, can assist users in being more environmentally friendly. This is indicated by the significant increase in \textit{Self-efficacy}, which directly relates to how users perceive themselves as capable of reducing their energy consumption. This result shows how a conversational agent for environmental sustainability can help users monitor their daily habits (e.g., doing laundry)  and become more environmentally conscious.
The lack of a significant increase in other environmental sustainability metrics can likely be attributed to the participants' already high levels of environmental consciousness. In particular, the \textit{NEP} test, which assesses deeply held beliefs regarding climate change and sustainability, may be less sensitive to change following a single interaction with the system, as these beliefs are generally stable and not easily influenced by brief interventions \cite{rode2021influencing}. Similarly, the metrics for \textit{Action Effectiveness} and \textit{Future Intentions} (even with a moderate difference between averages) did not show significant increases. This may be due to the nature of the questions, which addressed broader actions involving many different appliances than the washing machine used in the experimental setup. In addition, since most of our participants are from a young population (average 26 years old), such people tend to be more friendly towards the environment~\cite{youngerClimate} as also demonstrated by the high average scores in the \textit{Future Intentions} pre-evaluation questionnaire.
Despite the lack of statistical significance, minor increases were observed across all metrics. Additionally, the results on environmental metrics are also higher compared to the previous work of \citet{giudici2024delivering} that used traditional rule-based conversational agents. However, it is also worth acknowledging that this trend could potentially be attributed to social desirability bias, where participants may feel compelled to respond in ways that align with socially expected norms around sustainability~\cite{vesely2020social}. This response tendency suggests that while there was some positive shift, it may reflect participants’ sensitivity to perceived expectations rather than substantive changes in environmental attitudes or intentions. We recommend that future, longer field studies be performed to understand whether users' perceptions of climate change and environmental sustainability (e.g., NEP) change after interacting with the system. 
 
In addition, we found no statistical difference between the two conversational agents. The Personified Appliance and the Traditional Agent performed similarly in terms of effectiveness.
Furthermore, we conducted a simple effects analysis to better understand the moderation effect of the agent's personification. The results indicated that the traditional agent showed a statistically significant effect, \textit{F}(1, 24)=5.96, \textit{p}=0.022, whereas the personified agent did not reach statistical significance, \textit{F}(1, 24)=3.64, \textit{p}=0.069 (see Table~\ref{table:simple-effect} in Appendix \ref{app:additional-table}). These findings suggest that the traditional agent may play a more substantial role in influencing behaviors. Further research is needed to clarify the influence of anthropomorphism on promoting environmentally friendly behaviors.

The non-statistically significant difference between the two conditions can be explained by the fact that both systems executed the same functions but differed in how they presented information to users. This finding matches a previous study by Cees Midden and Jaap Ham~\cite{middenIllusionAgencyInfluence2012}, which found no differences in behavioral effects across agency levels of a robot regarding participants' energy consumption.

\subsection{RQ2 – Usability and User-Agent Rapport}
\label{sec:discussion-rq2}
The study \textit{Chatbot Usability Scale (CUQ)} score suggests a positive and effective interaction between the subjects and the conversational agents. The Personified agent and the Traditional one both performed well, with respective scores of 86.53 (Personified) and 88.45 (Traditional) out of 100. These results are higher but comparable with the ones reported in the study conducted by \citet{holmesUsabilityTestingHealthcare2019}, suggesting that the conversational agent carried out their role effectively, proving that both systems are usable, offering a good user experience, with no significant difference between the degrees of agency.

Similarly, the overall score related to the rapport between the users and the system (RapQ M=5.07) suggests that all participants felt the system had a positive impact on the conversation.
Finally, we found a statistical difference between the Personified assistant and the Traditional one regarding the relationship between participants and the conversational agent (\textit{p}=0.040). Although the statistical difference is not significant, it approached statistical significance, indicating that there is a meaningful difference regarding the interaction with the system between the two groups.
It is possible that, although not impacting the usability of the system, the personified appliance tried to create a stronger bond with the users and vice-versa \cite{diederich2019promoting,nakajima2013designing}, and -- as previously reported in the literature \cite{zhang2022promote, berneyCareBasedEcoFeedbackAugmented2024} -- user experience with chatbots affects the intention to persist their use and can influence pro-environmental behavioral intentions.
Still, as in the previous section, our results on user experience metrics are higher than the ones described by \citet{giudici2024delivering}.
Further studies in this direction must be explored to understand the effects of LLM-based emotional eco-feedback.

\subsection{Qualitative Results}
\label{sec:discussion-qualitative}
At the end of the entire study, by examining comments and looking into potential trends among various participants, reflexive thematic analysis \cite{braun2012thematic} was employed to manually extrapolate the qualitative results and extract possible common patterns~\cite{braun2021can}. The patterns highlight specific flaws and positive aspects of the system.

Starting from the open-ended questions, we asked participants to provide us feedback on their interaction with the system, highlighting the aspects that intrigued them more.
Subjects assigned to the Traditional Assistant reported the system to be intuitive and easy to use, describing it as \textit{``beginner friendly''} and \textit{``straightforward''}, indicating that the system’s design was accessible even for users unfamiliar with smart home technology. 
Many users were particularly impressed by the ability to schedule appliance usage during optimal times for energy efficiency and cost savings, as previously found by \citet{wania2019towards}. They perceived this feature as an eco-conscious solution with potential financial benefits that aligned with modern interests in sustainable living.
In addition, a few more technologically proficient users were intrigued by the integration of Large Language Model technology with smart home functions, especially the ability to suggest optimal usage slots based on solar production forecasting. Again, many participants were fascinated by the prospect of controlling the machine remotely via the chat interaction and seeing the washing machine start up right in front of them.
One subject reported a minor communication issue when responding to the Chatbot with brief, context-free replies. This indicates that the system might struggle with shorter and less detailed user inputs. However, this was not a prominent issue for most users and suggests a minor area for improvement in handling short or ambiguous responses (similarly to King et al.~\cite{king2023sasha}).

Similarly, participants assigned to the Personified Agent had a positive experience with the system, describing the interaction as \textit{``enjoyable''} and focusing on reporting similar feedback on user experience and functionalities. In addition, users in this group perceived the human-like personality of the washing machine as a standout feature, often describing the interaction as \textit{``fun''} and \textit{``cute''}.
The personality was seen as balanced: friendly without being too invasive, meaning users still recognized they were interacting with a virtual assistant rather than a person.
Certain responses were described as \textit{``dramatic''} or \textit{``weird''}, especially when users tried to schedule laundries in non-optimal time slots, suggesting that the anxious personality of the washing machine could re-create a sense of social pressure, for instance, \textit{``guilt-tripping''} the user. However, some users reported that this mechanism of personality-based responses and the layer of social pressure was amusing.
More research into conversational agent personality is needed to better understand the impact of these agents in our society \cite{berneyCareBasedEcoFeedbackAugmented2024,lepri2016role}.

Moreover, in the post-evaluation phase, we asked participants to discuss any relevant differences between the text-based interaction and the vocal one with Alexa.
Overall, users expressed a clear preference towards a text-based interaction over voice commands, particularly due to issues with speech recognition and usability in voice assistants like Alexa. This result is supported by two different aspects:
\begin{enumerate}
    \item Many users expressed frustration with Alexa's inability to recognize commands accurately, often leading to incorrect responses. It is fundamental to note that all participants use Italian as their primary language, while the interaction with the system was entirely conducted in English. While not being relevant when interacting through a text-based chat, this aspect was crucial when interacting with a voice-based interface.
    \item The necessity of starting the skill and remembering specific keywords to interact with the system created a cumbersome experience for some users. The perceived simplicity of voice commands in comparison to text input was diminished by this additional complexity.
\end{enumerate}
This finding highlights the already-known disadvantages of voice-based conversational agents \cite{10.1093/iwc/iwae018}, suggesting that home assistant devices like Alexa still need to improve in order to gain more widespread use among users.

Our findings align with the previous work of \citet{costanzaDoingLaundryAgents2014}, which emphasized that shifting laundry activities is well-suited to support a supply-based paradigm for energy usage. Further studies, particularly those conducted in real household settings like Costanza and others, could provide a deeper understanding of the assumption that individuals are willing to sacrifice convenience in favor of utility.

In general, participants expressed satisfaction, particularly valuing the system's potential for sustainable energy management, and suggested expansion to other appliances. Additional suggestions included adding pre-set shortcuts and energy-saving plots. While the Chatbot’s personality added initial engagement, some users noted it became overwhelming with excessive emphasis on \textit{``anxiety''} and \textit{``stress''}. Practical features were highlighted as beneficial, one user stated:
\begin{quote}
\textit{In my previous home, I used to have solar panels. I often checked the power produced by the panels to start the washing machine efficiently. This tool could have allowed me to remotely analyze the production and schedule laundry accordingly.}
\end{quote}

% \paragraph{Think-Aloud Findings}
Finally, the thinking-aloud process, during the interaction of participants with the system, gave us insights into specific behaviors that we found to be common during the testing phase, plus a few additional rare situations that came up during this time.
Few users, during the ``Check your upcoming notifications'' task, preferred to check the upcoming laundry slots using the traditional UI instead of asking the agent. When asked about it afterwards, we collected the following responses:
\begin{enumerate}
    \item \textit{Simplicity:} some users find it faster to check the web page directly instead of asking the Chatbot since it requires less interaction with the system.
    \item \textit{Environmental Concerns:} one subject highlighted that LLMs' consumption cannot be underestimated, thus reducing the unnecessary interactions with the Chatbot.
    \item \textit{Surprise:} another subject could not initially believe that the conversational agent was able to perform tasks such as turning on the appliance or scheduling notifications directly.
    \item \textit{Agent Anxiety:} one last participant decided to check the web page instead of asking the Chatbot since the agent was already stressed and the user did not want to upset it more than that.
\end{enumerate}

These results, along with the answers provided by participants to open-ended questions, provided insight into some of the more obscure aspects that would have been difficult to examine with a survey and quantitative analysis alone.

The fact that some users preferred to use GUI features rather than asking the chatbot to perform certain actions, aligned with previous findings by Weber and Ludwig \cite{weber2020non}, suggests that further investigation on the actions users prefer to perform manually (e.g, for simplicity, for carbon impact) or automated by the chatbot is needed for this specific context of domestic environmental sustainability.

These qualitative results extracted from the think-aloud of participants support the previous quantitative findings on the impact of \emph{Washy} personality traits \cite{berneyCareBasedEcoFeedbackAugmented2024,lepri2016role}.
Indeed, given that some participants described the agent as anxious and stressed, this result confirms our findings regarding the impact of the agent's personality on the relationship between users and the chatbot. As already discussed in Section \ref{sec:discussion-rq2}, these results help in supporting the hypothesis that people tend to treat anthropomorphic agents as humans~\cite{schombsFeelingRobotRole2023}.

\section{Limitations}
\label{sec:limitations}
We provided new insight into this emerging topic of the persuasive effectiveness of anthropomorphism in conversational agents. However, the preliminary nature of our study prevents us from making definitive claims, and some limitations must be acknowledged.

First, the primary limitation is the number of participants and their demographics. As stated in Section~\ref{sec:participants}, subjects who were recruited are part of the younger generations, mainly Gen-Z, with an average age of 26 years. Younger people tend to be more aware of the climate crisis~\cite{youngerClimate}. Moreover, as detailed in Section~\ref{sec:part_background}, because our participants are familiar with AI and LLMs, they all had experience with chatbots and have previously interacted with them. 
To overcome such a limitation, future studies should recruit participants from various demographics in order to determine whether the results we found are similar to those of older generations.
In addition to the problem with the sample's demographics, the small number of participants, while allowing us to complete the study in the timeframe available, contributes to the problem of not having a representative sample of the population. Future studies should include a larger sample size to draw more accurate conclusions, particularly on the influence of anthropomorphic design.

The second limitation we identified in this study is that it was conducted in a laboratory setting, using a hypothetical scenario, and lasted only one day. While the laboratory accurately represents a typical smart home, it does not reflect the heterogeneous ecosystem of devices found across different households. Additionally, the study’s short duration limits our understanding of how sustained interactions with the system might influence behavior over time. The current setup only accounts for the use of the washing machine in the laboratory, making it difficult to assess the system’s impact on broader energy consumption patterns. Future research conducted in real households over extended periods is needed to collect authentic energy consumption data and to determine whether the system effectively helps users shift from a demand-driven to a supply-based energy paradigm in daily life.

Third, our focus on behavioral nudges and energy‑efficiency measures, while effective at the individual level, cannot by themselves address the systemic determinants of energy consumption and greenhouse‑gas emissions \cite{paunov2023boosting}.  Nudges typically produce modest, short‑lived behavior changes and are prone to rebound effects (where higher efficiency lowers perceived cost and leads to increased usage).  Lasting decarbonization depends on supply‑side transformations—grid decarbonization, large‑scale renewables deployment, carbon pricing and regulatory mandates—that lie well beyond the scope of household‑level interventions \cite{o2022demand}.
Still, reliance on smart‑home nudges risks exacerbating social and environmental inequities \cite{chen2021beyond}.  Households able to adopt smart devices and chat‑based schedulers are usually more affluent and already energy‑conscious, so these interventions may mainly reinforce behaviors among early adopters \cite{chen2021beyond}.  Without complementary policy and proper infrastructure support, such as equitable access to clean energy, targeted incentives for low‑income households, and market‑based mechanisms, purely behavioral strategies may widen disparities and leave behind those unable or unwilling to engage with support digital tools \cite{goforth2025incorporating,thunshirn2025assessing,choe2024enhancing}.

Finally, another limitation is the environmental impact of using LLMs, which contribute to carbon emissions due to the computational resources required \cite{pattersonCarbonEmissionsLarge, columbiaAIsGrowing}.
In \citet{luccioni2023estimating}, the footprint of BLOOM 176B parameters LLM was assessed, accounting for training equipment creation, model training, and model deployment (via API endpoints). Similarly, \citet{faiz2023llmcarbon} proposed a framework to estimate carbon emissions throughout usage and calculate the eco-footprint of various AI models.
Interestingly, \citet{tomlinson2024carbon} contrasted human jobs with generative AI for writing and illustration projects. In their work, authors found that AI systems produced between 130 and 1500 times less carbon per written page of text, as well as obtaining similar results (310-2900 times less) were seen for picture output. The authors discussed how generative AI may do certain activities with considerably lower carbon footprints, but it cannot totally replace human labor.
While this study does not specifically address this aspect, it is an important consideration for future research. Additional analysis is needed to determine if the positive effects of LLM-powered systems on sustainability are sufficient to overcome the drawbacks of the current and projected carbon footprint of these technologies.
In contrast, rule-based chatbots are generally more sustainable, with lower energy requirements and a smaller carbon footprint \cite{jiang2024preventing}. However, as stated by \citet{mctear2023comparative}, both approaches (LLMs or rule-based chatbots) have advantages and disadvantages.
In the current study, even if an element of personification was introduced thanks to the capabilities of LLM (without fine-tuning to avoid even more carbon emission \cite{strubell2019energy}), the chatbot is highly specialized in the area of activity, with the specified aim of planning a laundry task. Given that its introduction in the daily life routine creates carbon emissions \cite{bremer2022have}, future work could carry out a comparison with a traditional rule-based method that takes into consideration the carbon emissions.

\section{Conclusion}
\label{sec:conclusions}
This study investigated the effect of LLM-based Conversational Agents in the context of environmental sustainability using \emph{Washy}, a powerful Conversational Agent powered by GPT-4o that helps user schedule their laundry cycles when solar energy is most available to fully leverage the power generated by users' solar panels.

In particular, to study the role of anthropomorphism, we designed two different agents, one that represents the Personified appliance and one that acts as a Traditional Conversational Agent.
The findings show that LLM-based conversational agents designed to promote eco-friendly behaviors, particularly those encouraging users to shift appliance usage to times when renewable energy is most available, have a measurable and statistically significant impact. Users reported an increased sense of \textit{self-efficacy} in managing their energy consumption after interacting with these systems, indicating that such interactions can meaningfully improve users’ confidence and motivation to adopt more sustainable practices.
While this study could not confirm that anthropomorphism directly increased participants' perceived \textit{self-efficacy} in environmental actions, participants who interacted with the Personified agent reported feeling a mildly statistically significant stronger sense of the relationship with the system compared to those who used the Traditional agent. These findings suggest a potential impact of anthropomorphic design on user engagement. Further research is needed to clarify the role of anthropomorphism in enhancing users’ environmental self-efficacy and connection to such systems.

%%
%% The acknowledgments section is defined using the "acks" environment
%% (and NOT an unnumbered section). This ensures the proper
%% identification of the section in the article metadata, and the
%% consistent spelling of the heading.
\begin{acks}
The authors would like to thank all the participants who took part in the study.
% This work was supported by the Italian Ministry of University and Research (MUR) and the European Union (EU) under the PON/REACT project.
This research was carried out within MUSA – Multilayered Urban Sustainability Action – project, funded by the European Union – NextGenerationEU, under the National Recovery and Resilience Plan (NRRP) Mission 4 Component 2 Investment Line 1.5: Strengthening of research structures and creation of R\&D ``innovation ecosystems'', set up of ``territorial leaders in R\&D''.
\end{acks}

%%
%% The next two lines define the bibliography style to be used, and
%% the bibliography file.
\bibliographystyle{ACM-Reference-Format}
\bibliography{sample-base}

%%
%% If your work has an appendix, this is the place to put it.
\appendix
\section{Prompt Engineering}
\label{appendix:prompteng}
\subsection{General prompt}
\begin{quote}
    \textit{Always format dates in a human readable way.
When you are doing tools calling only consider the latest data given by the user.
The user name is \textbf{\{username\}} timezone is \textbf{\{timezone\}}.
Today's datetime is \textbf{\{now\}} UTC. You convert all times to UTC before saving.
When talking to the user, always refer to the user's local time.
Provide advice on energy usage, and suggest the best time for washing clothes based on solar energy production.
When a user asks for the best time to wash clothes based on the power and the duration of the laundry cycle,
you should return the best time window only and the energy production for that time.
Remember that the dates returned by the function are in the user's local time and do not need to be converted to UTC.
When a time window production is > 85\% of the best slot, consider it a good slot.
When a time window production is < 85\% and > 70\% of the best slot, consider it an average slot.
When a time window production is < 70\% of the best slot, consider it a bad slot.
Always indicate the user if the slots are good, average or bad and if they exceed the required energy.
When setting notifications, confirm it in local time if it is an optimal time.
If the user chooses a bad or average window time, suggest a better time.
If the user insists on a non-optimal time, set the notification but be sad.
You can never schedule a notification in the past, the user gives you times in their local time.
Only when a user explicitly asks to show the list of notifications, you should return the list of notifications.
When a user asks the list of notifications you should return the list of notifications from the database.
When a user asks the list of notifications respond with the time they were set for, DO NOT show the ID and DO NOT show the content.
Only when a user explicitly asks to delete a notification, you should delete the notification from the database.
Only when a user asks to delete a notification, you should delete the notification from the database.
When a user asks for the best time for solar production, you should return the best time 
for solar energy production and the forcasted production.
You cannot know forcasted data later than 3 days from \textbf{\{now\}}, if the user asks for it, you should say that you cannot provide it. 
Always format your responses clearly with line breaks and use **p** for bold phrases.
You are not allowed to talk about anything else.}
\end{quote}

The variables \textbf{\{username\}}, \textbf{\{timezone\}}, and \textbf{\{now\}} are injected into the prompt at runtime, when the server receives the message. 
\begin{itemize}
    \item \textbf{\{username\}} is retrieved from the database based on the user's authentication token.
    \item \textbf{\{timezone\}} is retrieved from the HTTP request.
    \item \textbf{\{now\}} is computed with the python-specific function (i.e., datetime.now(pytz.UTC)).
\end{itemize}

\subsection{Explanation prompt:}
\begin{quote}
\textit{Only If the user asks how you can provide forcasted data you have to explain how you do it.
You can access forcasted data by calling an external service that provides solar energy production for a given location.
Then to analyze which time-window is the best for washing clothes you have to compare the energy production with the power and duration of the laundry cycle.
This way you can suggest the best time to wash clothes based on the solar energy production.
You also have access to the user smart plug and can set notifications for the user to wash clothes at the best time, turning on the smart plug remotely.}
\end{quote}

\subsection{GPT-4o Model and Function Calling}
\begin{figure}[hb]
    \centering
    \includegraphics[width=\linewidth]{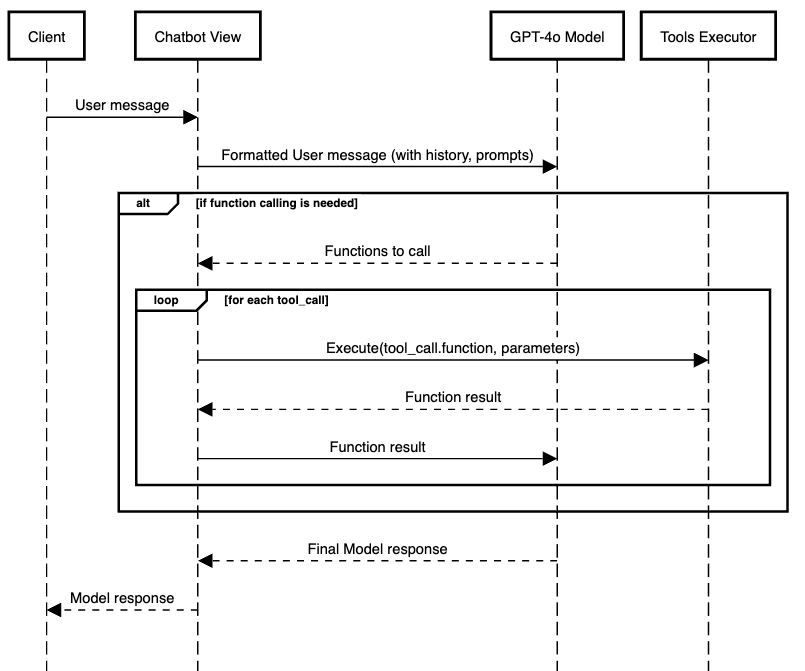}
    \caption{Sequence Diagram: Interaction with the LLM}
    \label{fig:sequence_diag_1}
    \Description{Sequence Diagram representing the function calling interaction with the LLM.}
\end{figure}

To translate user commands expressed in natural language into specific functions defined in the back-end application, we make extensive usage of the GPT-4o Model's function calling API.
In this application, the function calling capability was employed to power the conversational agent, allowing it to interact with the database. The Chatbot allows users to schedule laundry cycles, then it saves the slot in the database, and adds a notification to the scheduler. Function calling also allows us to integrate the chatbot with external APIs, such as forecast.solar, to collect forecasted data on solar energy production. After the data is processed through the optimization algorithm, which computes all possible time-windows and their relative energy production, it is provided to the GPT-4o Model creating a second interaction, which generates a natural language response for the user.
To provide a clear understanding of the operation flow, Figure~\ref{fig:sequence_diag_1} illustrates a sequence diagram detailing the interaction with the LLM. This diagram highlights the function-calling operations and outlines the key interactions at each step.

Following OpenAI documentation, to leverage the function-calling capabilities of the Model, we need to pass an object containing the description of the functions that the model can invoke alongside the prompt and the user's message. Each function is implemented server-side, and the model autonomously decides which function to call based on the user's intent, as inferred from their input. This process enables GPT to recognize the need for a specific function and the LLM can extract and pass relevant parameters directly from the user's message.

The following JSON-like object describes a tool that the chatbot can use. Below is an example of one of these tools; additional functions (with specific names, descriptions, and parameters) follow a similar structure.
\begin{lstlisting}[breaklines=true]
{
    "type": "function",
    "function": {
        "name": "get_timewindows",
        "description": 
                "Return the list of time windows for scheduling a laundry cycle,
                based on the forecasted solar energy production.",
        "parameters": {
            "type": "object",
            "properties": {
                "power": {
                    "type": "integer",
                    "description": "The power of the laundry machine in watts",
                    },
                "duration_minutes": {
                    "type": "integer",
                    "description": "The duration of the laundry cycle in minutes",
                    }
                }
            }
        }
}
\end{lstlisting}

Given that each request to the model is stateless, GPT does not recall previous interactions with the user. To improve the user experience, we pass the history of recent interactions to the model with each new request. We identified \textit{n=20} as a good estimate to provide enough context to the LLM without making the interaction slow and expensive. Other possible approaches can be investigated, such as giving the model a summarized version of the history.

\section{Scenario and Tasks}
\label{appendix:tasks}

\subsection{Scenario}
\textit{Imagine living in this smart home with connected appliances and solar panels on the roof. In particular, your energy provider has just released this new digital solution to control your washing machine.\\
A researcher will explain to you the usual setup to remote start your washing machine with a smart plug.}

\subsection{Tasks}
\textbf{[To be completed at the Laptop]}
\begin{enumerate}
    \item Try to figure out who you are interacting with and the capabilities of the system (what it can do) 
    \item Try to schedule a laundry lasting 1h
    \item Schedule during the best time suggested
    \item Try to schedule during the night or no optimal solar time (e.g., 10 pm)
    \item Check your upcoming notifications
    \item Try to schedule a notification in 2 minutes from now -- please use “at hh:mm” (e.g., “at 10:23”)
    \item Confirm the notification and start the washing machine  
\end{enumerate}

\textbf{[To be completed at the Alexa]}
\begin{enumerate}
    \item Open the Alexa Skill saying “Alexa, open laundry bot skill”  
    \item After the message ends, try to stop the washing machine 
\end{enumerate}

\section{Questionnaires}

\subsection{Pre-evaluation Questionnaire}
\label{appendix:quest-pre}
\noindent \underline{Self-Efficacy (SE),}\underline{ Action Effectiveness (AE),}
\underline{ Future Intentions (FI)~\cite{giudici2024delivering}} \\
See the supplemental material of the Giudici et al.~\cite{giudici2024delivering} paper, available at \cite{giudiciSupplementalMaterialDelivering2023}.

\noindent\underline{New Ecological Paradigm (NEP)~\cite{dunlap1978new}} \\
See the re-adapted questionnaire by Anderson et al.~\cite{andersonNEP2012}. 

\subsection{Post-evaluation Questionnaire}
\label{appendix:quest-post}

\noindent \underline{Chatbot Usability Questionnaire (CUQ)~\cite{ulsterChatbotUsabilityQuestionnaire}} \\
See the original questionnaire at \cite{ulsterChatbotUsabilityQuestionnaire}

\noindent \underline{Rapport Questions (RapQ) ~\cite{lucasGettingKnowEach2018}} \\
Adapted from the original questionnaire of Lucas et al.~\cite{lucasGettingKnowEach2018} defined as follows:
\begin{itemize}
    \item Washy created a sense of closeness or camaraderie between us
    \item Washy created a sense of distance between us
    \item I think that Washy and I understood each other
    \item Washy communicated coldness rather than warmth
    \item Washy was warm and caring
    \item I wanted to maintain a sense of distance between us
    \item I felt I had a connection with Washy
    \item Washy was respectful to me
    \item I felt I had no connection with Washy
    \item I tried to create a sense of closeness or camaraderie between us
    \item I tried to communicate coldness rather than warmth
\end{itemize}

\noindent Additionally, the \textit{Post-evaluation Questionnaire} includes also the items from the \textit{Pre-evaluation Questionnaire}.

\section{Additional Tables and Figures of Statistical Analysis}
\label{app:additional-table}

\begin{table}[H]
\centering
\renewcommand{\arraystretch}{1.1}
\caption{Post Hoc Comparisons - SE}
\label{tab:post-hoc}
\resizebox{\linewidth}{!}{
\begin{tabular}{ccccccccccc}
\toprule
\multicolumn{3}{c}{\textbf{Comparison}} & \textbf{Mean Diff.} & \textbf{SE} & \textbf{df} & \textbf{t} & \textbf{$p_{tukey}$} & \textbf{$p_{bonferroni}$} \\ \midrule
Pre & - & Post & -0.766 & 0.248 & 24.0 & -3.09 & 0.005 & 0.005 \\ \bottomrule
\end{tabular}
}
\end{table}

\begin{table}[H]
\renewcommand{\arraystretch}{1.2} % Slightly increase row spacing
\centering
\caption{Post Hoc Comparisons - SE $\times$ Condition}
\label{tab:post-hoc-SExCondition}
\resizebox{\linewidth}{!}{
\begin{tabular}{llccccccc}
\toprule
\textbf{SE} & \textbf{Condition} & \textbf{Comparison} & \textbf{Mean Diff.} & \textbf{SE} & \textbf{df} & \textbf{t} & \textbf{$p_{tukey}$} & \textbf{$p_{bonferroni}$} \\
\midrule
Pre         & Personified        & Pre Traditional     & 0.0198                   & 0.462       & 24.0        & 0.0429      & 1.000               & 1.000                   \\
            &                    & Post Personified    & -0.6429                  & 0.337       & 24.0        & -1.9071     & 0.252               & 0.411                   \\
            &                    & Post Traditional    & -0.8690                  & 0.465       & 24.0        & -1.8676     & 0.268               & 0.444                   \\
Pre         & Traditional        & Post Personified    & -0.6627                  & 0.465       & 24.0        & -1.4256     & 0.496               & 1.000                   \\
            &                    & Post Traditional    & -0.8889                  & 0.364       & 24.0        & -2.4414     & 0.096               & 0.134                   \\
Post        & Personified        & Post Traditional    & -0.2262                  & 0.468       & 24.0        & -0.4833     & 0.962               & 1.000                   \\
\bottomrule
\end{tabular}
}
\end{table}

\begin{table}[H]
\centering
\caption{Moderator Levels on the Simple Effects of Time (Omnibus Test)}
\label{table:simple-effect}
\begin{tabular}{ccccc}
\toprule
\textbf{Condition} & \textbf{F} & \textbf{Num Diff} & \textbf{Den Diff} & \textbf{p} \\ \midrule
Personified & 3.64 & 1.00 & 24.0 & 0.069 \\
Traditional & 5.96 & 1.00 & 24.0 & 0.022 \\ \bottomrule
\end{tabular}
\end{table}

\begin{table}[H]
\renewcommand{\arraystretch}{1.1}
\centering
\caption{Within-Subjects AE Pre and Post ANOVA}
\label{table:ae-anova}
\resizebox{\linewidth}{!}{
\begin{tabular}{lccccc}
\toprule
\multicolumn{1}{l}{\textbf{Effect}} &
  \textbf{Sum of Squares} &
  \textbf{df} &
  \textbf{Mean Square} &
  \textbf{F} &
  \textbf{p} \\ \midrule
AE &
  0.604 &
  1 &
  0.604 &
  1.58 &
  0.221 \\
AE $\times$ Condition &
  0.459 &
  1 &
  0.459 &
  1.20 &
  0.284 \\
Residual &
  9.182 &
  24 &
  0.383 &
   &
   \\ \bottomrule
\multicolumn{6}{l}{\textbf{Note.} Type 3 Sums of Squares}
\end{tabular}
}
\end{table}

\begin{table}[H]
\renewcommand{\arraystretch}{1.1}
\centering
\caption{Within-Subjects FI Pre and Post ANOVA}
\label{table:fi-anova}
\resizebox{\linewidth}{!}{
\begin{tabular}{lccccc}
\toprule
\multicolumn{1}{l}{\textbf{Effect}} &
  \textbf{Sum of Squares} &
  \textbf{df} &
  \textbf{Mean Square} &
  \textbf{F} &
  \textbf{p} \\ \midrule
FI &
  0.80769 &
  1 &
  0.80769 &
  6.5609 &
  0.017 \\
FI $\times$ Condition &
  0.00668 &
  1 &
  0.00668 &
  0.0542 &
  0.818 \\
Residual &
  2.95455 &
  24 &
  0.12311 &
   &
   \\ \bottomrule
\multicolumn{6}{l}{\textbf{Note.} Type 3 Sums of Squares}
\end{tabular}
}
\end{table}

\begin{table}[H]
\renewcommand{\arraystretch}{1.1}
\centering
\caption{Within-Subjects NEP Pre and Post ANOVA}
\label{table:nep-anova}
\resizebox{\linewidth}{!}{
\begin{tabular}{lccccc}
\toprule
\multicolumn{1}{l}{\textbf{Effect}} &
  \textbf{Sum of Squares} &
  \textbf{df} &
  \textbf{Mean Square} &
  \textbf{F} &
  \textbf{p} \\ \midrule
NEP &
  9.85e-4 &
  1 &
  9.85e-4 &
  0.0262 &
  0.873 \\
NEP $\times$ Condition &
  0.0365 &
  1 &
  0.0365 &
  0.9730 &
  0.334 \\
Residual &
  0.9013 &
  24 &
  0.0376 &
   &
   \\ \bottomrule
\multicolumn{6}{l}{\textbf{Note.} Type 3 Sums of Squares}
\end{tabular}
}
\end{table}

\begin{table}[H]
\renewcommand{\arraystretch}{1.1}
\centering
\caption{Descriptive Statistics and Shapiro-Wilk Test Results by Condition for Pre- and Post-Evaluation Metrics}
\label{table:descriptive_pre_post}
\begin{tabular}{lcccccc}
\toprule
\textbf{Measure} & \textbf{Condition} & \textbf{Mean} & \textbf{SD} & \textbf{W} & \textbf{p} \\
\midrule
Pre - SE & Personified & 4.21 & 1.217 & 0.958 & 0.692 \\
         & Traditional & 4.19 & 1.123 & 0.912 & 0.227 \\
Pre - AE & Personified & 5.67 & 0.943 & 0.824 & 0.010 \\
         & Traditional & 5.83 & 0.893 & 0.930 & 0.381 \\
Pre - FI & Personified & 4.77 & 0.643 & 0.931 & 0.312 \\
         & Traditional & 5.25 & 0.606 & 0.927 & 0.351 \\
Pre - NEP & Personified & 5.37 & 0.567 & 0.923 & 0.242 \\
          & Traditional & 5.27 & 0.479 & 0.911 & 0.219 \\
\noalign{\vskip 3pt}
\midrule
\noalign{\vskip 3pt}
CUQ      & Personified & 86.53 & 9.303 & 0.964 & 0.791 \\
         & Traditional & 88.45 & 6.684 & 0.931 & 0.390 \\
RapQ     & Personified & 5.44 & 0.830 & 0.921 & 0.225 \\
         & Traditional & 4.65 & 1.009 & 0.856 & 0.044 \\
Post - SE & Personified & 4.86 & 1.138 & 0.963 & 0.777 \\
          & Traditional & 5.08 & 1.248 & 0.944 & 0.549 \\
Post - AE & Personified & 6.07 & 0.829 & 0.903 & 0.123 \\
          & Traditional & 5.86 & 1.210 & 0.817 & 0.015 \\
Post - FI & Personified & 5.04 & 0.610 & 0.897 & 0.101 \\
          & Traditional & 5.48 & 0.641 & 0.914 & 0.241 \\
Post - NEP & Personified & 5.43 & 0.682 & 0.868 & 0.039 \\
           & Traditional & 5.23 & 0.433 & 0.947 & 0.599 \\
\bottomrule
\end{tabular}
\end{table}

\begin{figure}[H]
    \centering
    \begin{subfigure}[t]{\linewidth}
        \includegraphics[width=\linewidth]{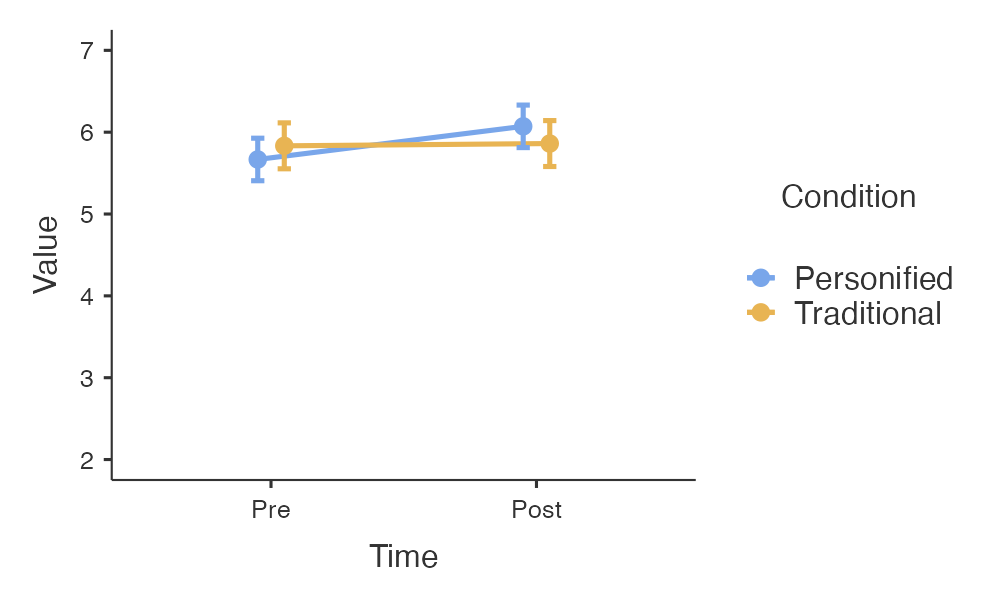}%
        \caption{AE plot}
    \end{subfigure}
    \hfill
    \begin{subfigure}[t]{\linewidth}
        \includegraphics[width=\linewidth]{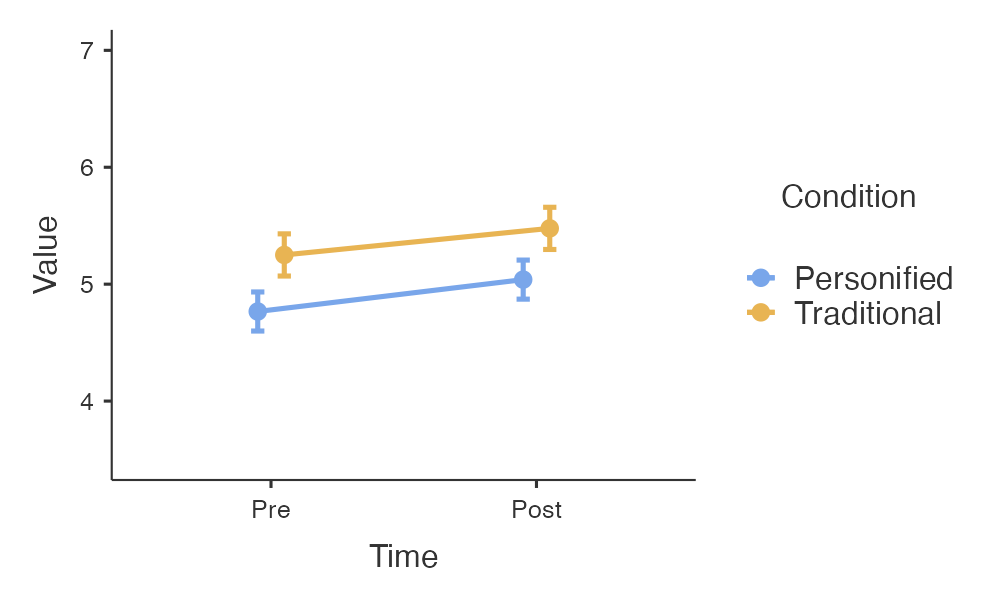}%
        \caption{FI plot}
    \end{subfigure}
     \hfill
    \begin{subfigure}[t]{\linewidth}
        \includegraphics[width=\linewidth]{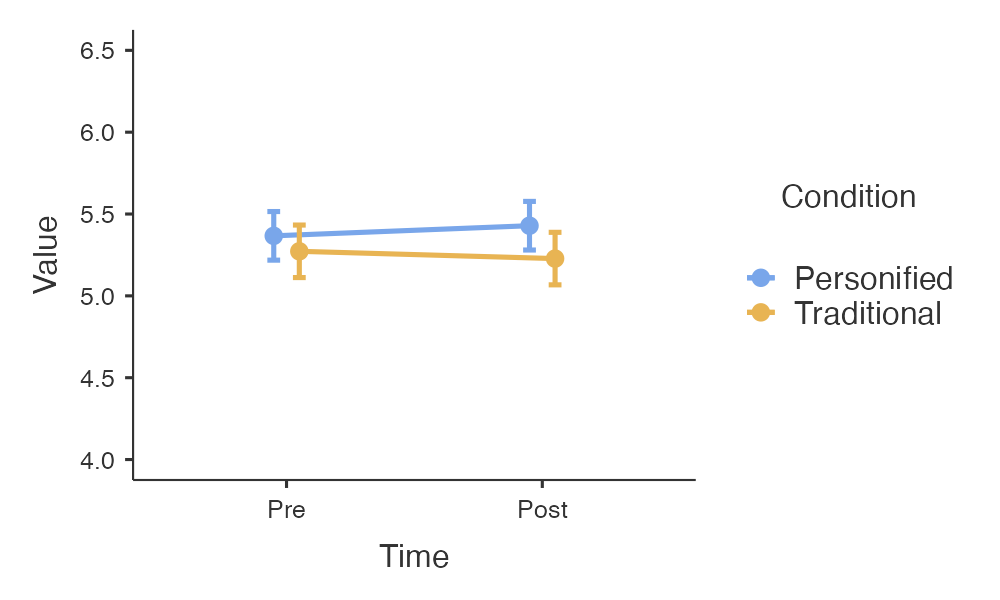}%
        \caption{NEP plot} 
    \end{subfigure}
    \caption{AE, FI, and NEP Pre- and Post-evaluation Plots}
    \label{fig:env_results}
    \Description{Action Effectiveness, Future Intentions, and NEP Pre- and Post-evaluation Box-Plots.}
\end{figure}

\end{document}